\documentclass[lettersize,journal]{IEEEtran}

\IEEEoverridecommandlockouts
\usepackage{cite}
\usepackage{amsmath,amssymb,amsfonts}
\usepackage{float}
\usepackage{graphicx}
\usepackage{textcomp}
\usepackage{xcolor}
\usepackage{array}
\usepackage{multicol}
\usepackage{multirow}
\allowdisplaybreaks
\usepackage{subcaption}
\usepackage[ruled,vlined,linesnumbered]{algorithm2e}
\usepackage{bm}
\usepackage{tabularx}
\usepackage{makecell}
\usepackage{soul}
\usepackage{caption} 
\newtheorem{proposition}{Proposition}

\def\BibTeX{{\rm B\kern-.05em{\sc i\kern-.025em b}\kern-.08em
    T\kern-.1667em\lower.7ex\hbox{E}\kern-.125emX}}
\begin{document}

\title{Deep Learning Based Superposition Coded Modulation for Hierarchical Semantic Communications over Broadcast Channels
\author{
Yufei~Bo, Shuo~Shao, Meixia~Tao}
\thanks{Y. Bo, and M. Tao are with the Department of Electronic Engineering and Shanghai Key Laboratory of Digital Media Processing and Transmission, Shanghai Jiao Tong University, Shanghai, 200240, China. (emails: \{boyufei01, mxtao\}@sjtu.edu.cn)}
\thanks{S. Shao is with the School of Cyber Science and Engineering, Shanghai Jiao Tong University, Shanghai, 200240, China. (email: shuoshao@sjtu.edu.cn)}
\thanks{
Part of this work was presented at IEEE GLOBECOM 2023 \cite{confVersion}.}
}

\maketitle

\begin{abstract}

We consider multi-user semantic communications over broadcast channels. While most existing works consider that each receiver requires either the same or independent semantic information, this paper explores the scenario where the semantic information desired by different receivers is different but correlated. In particular, we investigate semantic communications over Gaussian broadcast channels where the transmitter has a common observable source but the receivers wish to recover hierarchical semantic information in adaptation to their channel conditions. Inspired by the capacity achieving property of superposition coding, we propose a deep learning based superposition coded modulation (DeepSCM) scheme.  Specifically, the hierarchical semantic information is first extracted and encoded into basic and enhanced feature vectors. A linear minimum mean square error (LMMSE) decorrelator is then developed to obtain a refinement from the enhanced features that is uncorrelated with the basic features. Finally, the basic features and their refinement are superposed for broadcasting after probabilistic modulation. Extensive experiments are conducted for two-receiver image semantic broadcasting with coarse and fine classification as hierarchical semantic tasks. DeepSCM outperforms the benchmarking coded-modulation scheme without a superposition structure as well as the classic separate source-channel coding baselines, especially with large channel disparity and high order modulation. It also approaches the performance upperbound as if there were only one receiver.

\end{abstract}


\section{Introduction}

Semantic communication has recently emerged to deliver intelligent services. 
Departing from the traditional focus on source recovery, it revolutionizes data transmission by extracting and transmitting the ``semantic information'' crucial for the intelligent tasks at the receiver\cite{survey:beyond, survey:view, survey:overview, survey:opportunity}.
Therefore, semantic communications can significantly improve transmission efficiency and service quality over traditional Shannon-type communications.

Leveraging the advantages of deep learning, semantic communication has mainly adopted neural networks (NNs) for semantic coding, which demonstrates superior performance in point-to-point communication scenarios. 
In particular, NN-based semantic coding has been used to substitute the conventional source coding and/or channel coding to transmit various source data types
including speeches\cite{semantic:speech}, texts\cite{semantic:text, semantic:text2, semantic:text3}, images\cite{semantic:image, semantic:image2, semantic:image3, gunduz2019jscc1}, videos\cite{semantic:video, semantic:video2} as well as multi-modal data\cite{mac}.
It also facilitates the execution of a wide range of intelligent semantic tasks, such as object detection\cite{objectdetection}, classification\cite{semantic:image}, and question-answering\cite{mac}.
Different types of NNs have been employed for the semantic coding of different source data types.
For example, the Resnet \cite{resnet} is often used for image sources\cite{semantic:image} while Transformer\cite{vaswani2017attention} and BERT\cite{devlin2018bert} is often used for text sources\cite{semantic:text, one2many}.
Moreover, while earlier semantic communication systems often employ analog modulation to directly transmit through the channel the real-valued output of the neural semantic encoder\cite{semantic:speech, semantic:image, semantic:text}, recent advancements have brought forth digital semantic communications by taking into account digital modulation explicitly in \cite{vectorquantization,xie2023robust,jaccq,jcm, bo2023joint}.
In particular, the works \cite{jcm, bo2023joint} introduce a novel joint coding-modulation framework for end-to-end design of digital semantic communication systems.
Therein, variational autoencoder (VAE) is applied to generate the transition probability from source data to discrete constellation symbols, which resolves the non-differentiable issue of digital modulation and learns the probabilistic shaping of constellations for the operating channel condition. 

\begin{table*}[t]
\small
\renewcommand{\arraystretch}{1.1}
  \begin{center}
    \caption{Summary of Notation.}
    \begin{tabularx}{\textwidth}{lXXX}
    \hline
      Notation & Description \\
      \hline
      $\mathbf{S}_1$, $\mathbf{\hat S}_1$ & Coarse-grained semantic information and its recovery at Receiver 1 \\
      $\mathbf{S}_2$, $\mathbf{\hat S}_2$ & Fine-grained semantic information and its recovery at Receiver 2 \\
      $\mathbf{X}$, $\mathbf{\hat X}_1$, $\mathbf{\hat X}_2$ & Observable image source and its recoveries respectively at Receiver 1 and Receiver 2 \\
      $\mathbf{U}_1$ & Basic EFV\\
      $\mathbf{U}_2$ & Enhanced EFV\\
      $\mathbf{R}$ & Successive refinement vector\\
      $\mathbf{Y}_1$, $\mathbf{Y}_2$ & Inner constellation sequence and outer constellation sequence\\
      $\mathbf{Y}$ & The super-constellation sequence to be sent into the channel\\ 
      $\mathbf{Z}_i$ & Received sequence of Receiver $i$\\
      $f_{\boldsymbol{\theta}_1}(\cdot)$ & Basic semantic encoder\\
      $f_{\boldsymbol{\theta}_2}(\cdot)$ & Enhancement semantic encoder\\
      $\mathbf{W}$, $\mathbf{b}$ & Weight and bias of the LMMSE decorrelator\\
      $\boldsymbol{\rho}_i$ & NN parameters of Modulator $i$\\
      $g_{\boldsymbol{\eta}_1}(\cdot)$, $g_{\boldsymbol{\eta}_2}(\cdot)$ & Semantic decoders for image recovery of Receiver 1 and Receiver 2\\
      $g_{\boldsymbol{\psi}_1}(\cdot)$, $g_{\boldsymbol{\psi}_2}(\cdot)$ & Semantic decoders for coarse classification of Receiver 1 and fine classification of Receiver 2\\
      $n$ & Number of channel uses\\
      $\alpha$ & Power allocation factor\\
      \hline
    \label{notations}
    \end{tabularx}
  \end{center}
  \vspace{-0.9cm}
\end{table*}

While substantial research has been devoted to point-to-point semantic communications\cite{semantic:speech,semantic:text,semantic:text2,semantic:text3,semantic:image,semantic:image2,semantic:image3,semantic:video,semantic:video2,jcm}, recently there is a growing interest in semantic communications over multi-user channels.
Many-to-one semantic communication systems with multiple transmitters and a single receiver are considered in \cite{distributedJSCC, yilmaz2023distributed,lin2023channel, mac,cooparative,qin:multiuser,mdma}.
Specifically, the work \cite{distributedJSCC} introduces a distributed joint source-channel coding (JSCC) scheme for correlated image sources, where each source is transmitted through a dedicated and independent noisy channel to the common receiver.
A distributed JSCC scheme for image transmission over Gaussian multiple access channel (MAC) is considered in \cite{yilmaz2023distributed}.
The work \cite{lin2023channel} considers a multi-user fading channel and introduces a channel-transferable semantic communication for orthogonal frequency division multiplexing with non-orthogonal multiple access (OFDM-NOMA) system.
Moreover, the works \cite{mac,cooparative,qin:multiuser} consider multi-user channels with a multi-antenna receiver, 
where \cite{cooparative} considers cooperative object identification, while \cite{mac} and \cite{qin:multiuser} address the transmission of multimodal data.
Additionally, a novel multiple access technology called model division multiple access (MDMA) is proposed in \cite{mdma}, where both uplink and downlink scenarios are considered.

Research on semantic communications over one-to-many broadcast channels has also been conducted\cite{ma2023features,wu2023fusion,li2023non,one2many,lu2023self}, which is aligned with the focus of this paper.
Specifically, the work \cite{ma2023features} introduces a feature-disentangled semantic broadcasting communication system. Therein, the extracted receiver-specific semantic features from a common source are encoded using conventional bit-level source and channel coding.  
The work \cite{wu2023fusion} proposes an image semantic fusion scheme that combines the semantic features for different receivers into a unified latent representation before broadcasting. 
The work \cite{li2023non} introduces a downlink NOMA-enhanced semantic communication system with diverse modalities of sources. Therein, the semantic features of different users are mapped into discrete constellations via an asymmetric quantizer and then superposed for broadcasting.
Moreover, to better deal with diverse channel conditions among multiple receivers, the authors in \cite{one2many} propose to first jointly train the transmitter with one receiver, then employ transfer learning for the other receiver. 
The work \cite{lu2023self} proposes a reinforcement learning based self-critical alternate learning algorithm for adapting to different channel conditions with sentence generation task.
Notably, these existing research efforts largely focus on the scenarios where the multiple receivers either require independent semantic information\cite{ma2023features,wu2023fusion,li2023non} or desire the common semantics\cite{one2many, lu2023self}.
It remains unexplored the scenarios where the required semantic information by different receivers is different but correlated.

In this paper, we investigate semantic communication over broadcast channels where the different but correlated semantic information needed by multiple receivers exhibits a hierarchical structure.
Specifically, the common observable source and the associated semantic information are modeled as a Markov chain.
The receivers with poor channel conditions wish to recover the observable source and the coarse-grained semantic information while the receivers with better channel conditions wish to recover the observable source and the finer-grained semantic information.
Our goal is to design an efficient digital semantic communication framework that can exploit such hierarchical structure of semantic information and accommodate the diverse channel conditions of different receivers.

To this end, we propose a novel deep learning-based superposition coded modulation (DeepSCM) scheme for hierarchical semantic communications over Gaussian degraded broadcast channels.
The main idea is to utilize NNs to separately extract and encode the different levels of semantic information from the common source, then combine them using supposition coding before broadcasting.  
Thereby, each receiver can decode its desired level of semantic information from the superposition-structured symbols according to its own channel condition.

Specifically, let us consider the two-user broadcast channel without loss of generality.
We first use two separate neural semantic encoders to map the common observable source respectively into a basic encoded feature vector (EFV) (for the poor receiver) and an enhanced EFV (for the good receiver).
Note that the two EFVs are generally correlated. 
As a classic capacity-achieving scheme for degraded broadcast channels\cite{cover:broadcast}, superposition coding, however, applies to independent data streams. 
We thus introduce a novel learning-based linear minimum mean square error (LMMSE) decorrelator module in our transmitter design. 
This decorrelator aims to extract the residual information from the enhanced EFV that is uncorrelated with the basic EFV, referred to as the successive refinement vector of the basic EFV.
Theoretical proof is further provided to establish that the successive refinement vector and the basic EFV can indeed be uncorrelated through optimization.
After that, we design a superposition modulation strategy, associating the basic EFV with the inner layer and the successive refinement vector with the outer layer. 
As a result, the poor receiver can decode the inner layer and recover the first level of semantic information.
At the same time, the good receiver can further decode the outer layer, therefore recovering the enhanced level of semantic information. 
Moreover, to ensure the convergence and the performance of the training process, we devise a three-stage training strategy for the training of DeepSCM according to the superposition structure.


\begin{figure*}[t]
\centering
\includegraphics[scale=0.6]{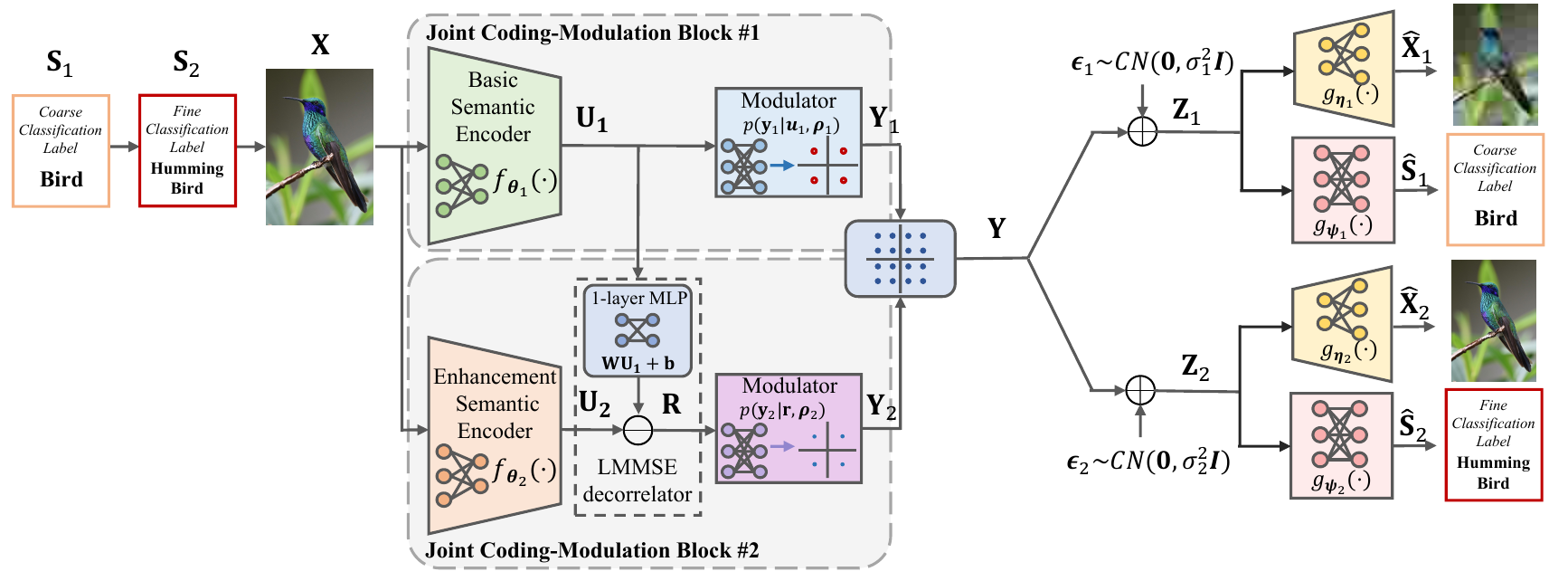}
\caption{Framework of the proposed DeepSCM scheme.}
\label{overview}
\vspace{-0.3cm}
\end{figure*} 


Extensive experiments on real-world image datasets have validated the performance advantages of the proposed DeepSCM across various modulation orders, compression rates, and channel disparity levels. 
It outperforms the conventional coded modulation scheme without a superposition structure, where the transmitter uses a single semantic encoder that is jointly trained with the semantic decoders at the two receivers.
This performance advantage grows as the channel signal-to-noise (SNR) gap between the two receivers widens and the modulation order increases.
Furthermore, the proposed DeepSCM scheme can simultaneously approach the performance upperbound for both receivers, as if serving only one receiver.


The rest of the paper is organized as follows.
The overall framework of DeepSCM is presented in Section II.
Section III describes the transmitter design in detail, including the main components of the transmitter and the training strategy.
In Section IV, we evaluate the performance of our proposed DeepSCM scheme through extensive experiments.
Finally, we conclude the paper in Section V.

Throughout this paper, we use upper-case letters ($X$) and lower-case letters ($x$) to respectively denote random variables and their realizations.
We use $h(X)$ to denote the differential entropy of the continuous variable $X$.
The statistical expectation of $X$ is denoted as $\mathbf{E}\lbrack X \rbrack$.
The covariance matrix of a vector random variable $\mathbf{P}$ is denoted as $\textit{Var}[\mathbf{P}]$, and $\textit{Cov}[\mathbf{P}, \mathbf{Q}]$ denotes the cross-covariance matrix of $\mathbf{P}$ and $\mathbf{Q}$.
We use $\mathcal{CN}(\mu, \sigma^2)$ to denote the circularly symmetric complex Gaussian distribution with mean $\mu$ and variance $\sigma^2$.
We use $\mathbf{I}_{k\times k}$ to denote the identity matrix with dimensions $k\times k$.
We use $[M]$ to represent the set $\{1, ..., M\}$ for $M\in \mathbb{Z}^+$, and occasionally use the short-hand Tx for a transmitter and Rx for a receiver.
All important notations used in this paper are summarized
in Table \ref{notations}.

\section{Overall Framework}

This paper mainly focuses on semantic communication over two-user broadcast channels. 
The extension to $K$ ($K>2$) users shall be briefly discussed at the end of this section.

Fig.~\ref{overview} illustrates the overall framework of the DeepSCM scheme over a two-user Gaussian degraded broadcast channel.
Without loss of generality, we assume that Receiver 2 has a higher channel SNR than Receiver 1. 
There is an observable source $\mathbf{X}$ associated with implicit hierarchical semantic information, namely the coarse-grained semantic information $\mathbf{S}_1$ and the fine-grained semantic information $\mathbf{S}_{2}$. 
Naturally, they form a Markov chain as
\begin{equation}
  \mathbf{S}_1\to\mathbf{S}_{2}\to\mathbf{X}. 
  \label{markov}
\end{equation}
Following the conventional setup of semantic communications\cite{image1, semantic:image3}, each receiver needs to recover both the semantic information and the observable source, which reflects the practical demands in many real-world scenarios where both humans and machines are involved in decision-making.
Receiver 1 requires the observable source $\mathbf{X}$ and the coarse-grained semantic information $\mathbf{S}_1$, the recovery of which we respectively denote as $\hat{\mathbf{X}}_1$ and $\hat{\mathbf{S}}_1$.
Meanwhile, as Receiver 2 has a larger channel capacity, it requires the observable source $\mathbf{X}$ and the fine-grained semantic information $\mathbf{S}_2$, the recovery of which we respectively denote as $\hat{\mathbf{X}}_2$ and $\hat{\mathbf{S}}_2$.

Specifically, in this paper we focus on image semantic communications for both image recovery and classification.
Namely, the observable source $\mathbf{X}\in\mathbb{R}^d$ is image data, where $d$ represents the dimension of the images.
The coarse-grained semantic information $\mathbf{S}_1\in\{1,2,...,L_1\}$ is the label of coarse image classification with $L_1$ classes.
Similarly, the fine-grained semantic information $\mathbf{S}_2\in\{1,2,...,L_2\}$ is the label of fine image classification with $L_2$ classes, where $L_2>L_1$.
This hierarchical structure of the observable source and the semantics can well model the real-world scenarios.
For instance, in wildlife monitoring, semantic information $\mathbf{S}_1$ can represent the general categories of animals, such as birds, bears, \emph{etc}, while semantic information $\mathbf{S}_2$ can represent the specific species within those categories, such as sparrows, cardinals, black bears, brown bears, \emph{etc}.
Besides image classification, the above Markov-chain based hierarchical semantic information model \eqref{markov} is also suitable for other tasks. 
For example, $\mathbf{S}_1$ could represent the location of prominent objects in an image relevant to the task of object detection, while $\mathbf{S}_2$ could indicate both the location and classification of the objects that are relevant to the task of image segmentation.

\begin{figure}
     \centering
     \begin{subfigure}[b]{0.24\textwidth}
         \centering
         \includegraphics[width=0.9\textwidth]{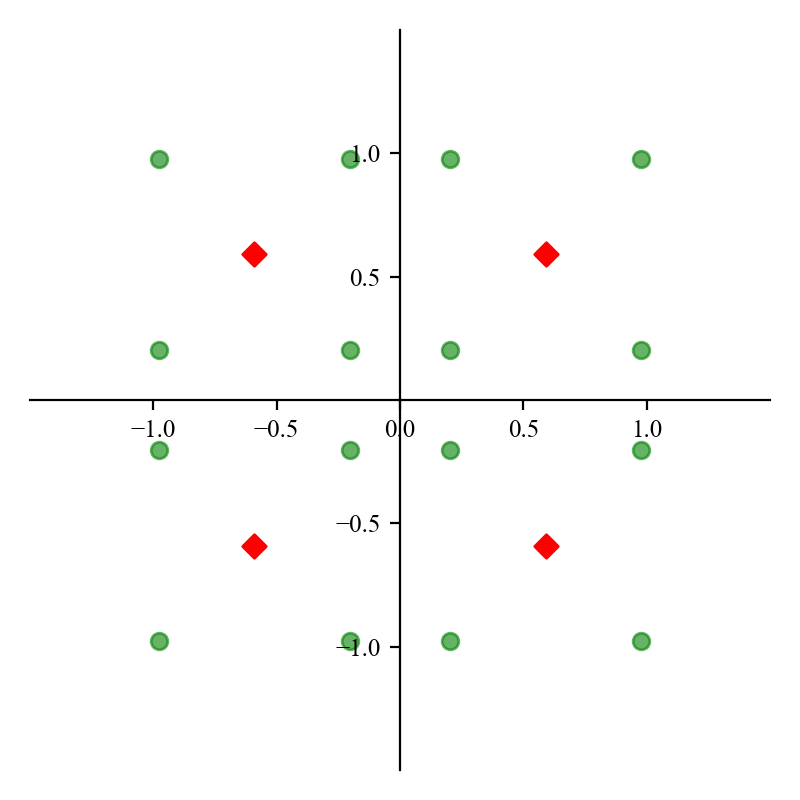}
         \caption{4QAM$\times$4QAM, $\alpha=0.7$.}
     \end{subfigure}
          \begin{subfigure}[b]{0.24\textwidth}
         \centering
         \includegraphics[width=0.9\textwidth]{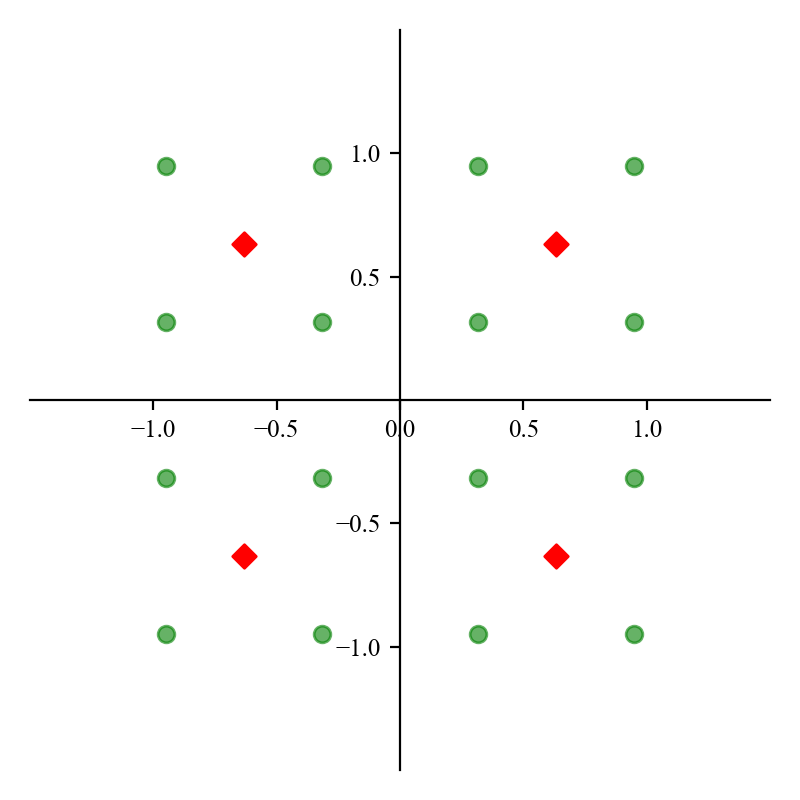}
         \caption{4QAM$\times$4QAM, $\alpha=0.8$.}
     \end{subfigure}
          \begin{subfigure}[b]{0.24\textwidth}
         \centering
         \includegraphics[width=0.9\textwidth]{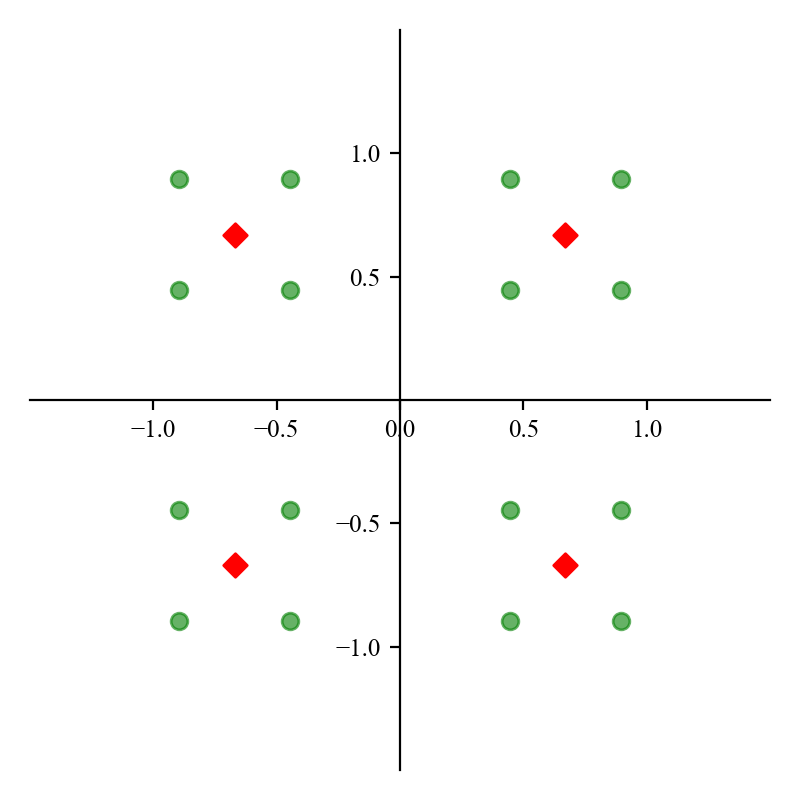}
         \caption{4QAM$\times$4QAM, $\alpha=0.9$.}
     \end{subfigure}
     \begin{subfigure}[b]{0.24\textwidth}
         \centering
         \includegraphics[width=0.9\textwidth]{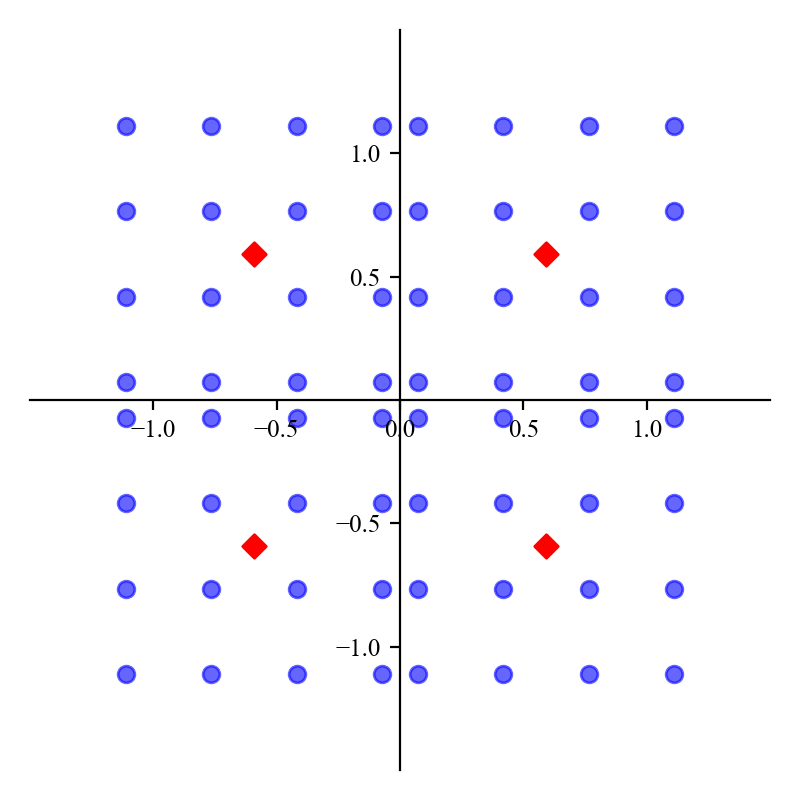}
         \caption{4QAM$\times$16QAM, $\alpha=0.7$.}
     \end{subfigure}
          \begin{subfigure}[b]{0.24\textwidth}
         \centering
         \includegraphics[width=0.9\textwidth]{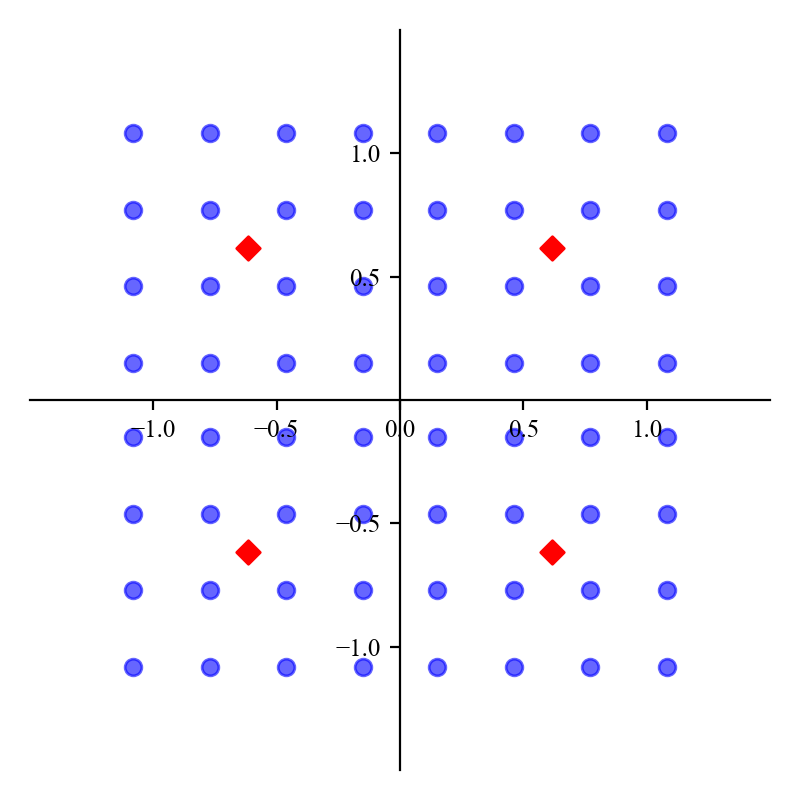}
         \caption{4QAM$\times$16QAM, $\alpha=0.76$.}
     \end{subfigure}
          \begin{subfigure}[b]{0.24\textwidth}
         \centering
         \includegraphics[width=0.9\textwidth]{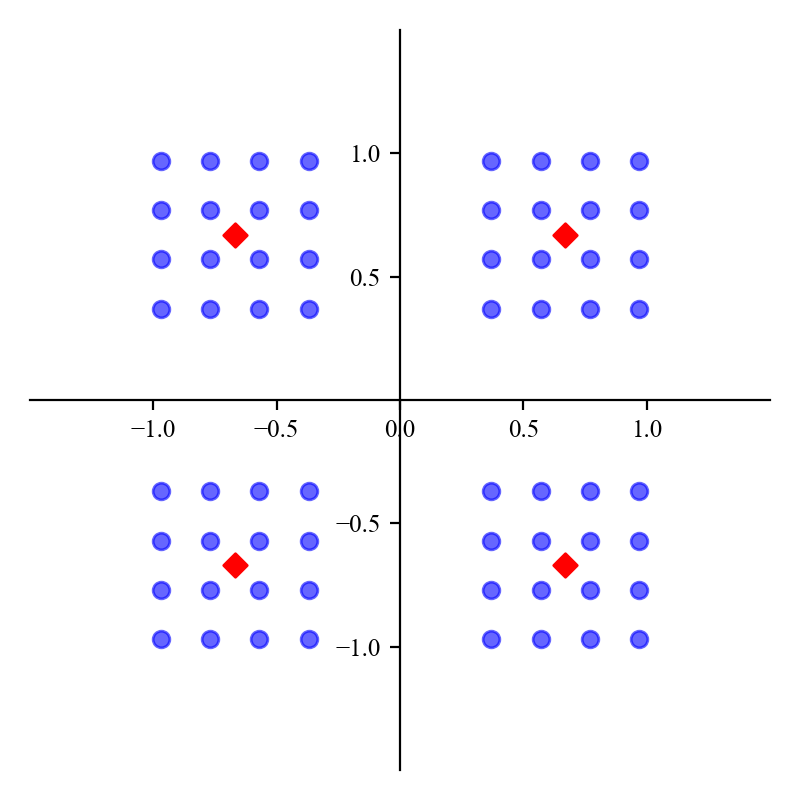}
         \caption{4QAM$\times$16QAM, $\alpha=0.9$.}
     \end{subfigure}
        \caption{Constellations after superposition with various values of $\alpha$. The red diamonds indicate the inner constellation multiplied by the PAF, and the blue dots indicate the resulted super-constellation. All symbols are assumed to be uniformly distributed in this figure with $P=1$.
        Notably, (b) and (e) present rectangular 16QAM and 64QAM, respectively.
        }
        \label{examples}
        \vspace{-0.4cm}
\end{figure}

The transmitter extracts semantic features in a hierarchical way, corresponding to the hierarchical sources intended for the two receivers.
At the transmitter, there are two joint coding-modulation (JCM) blocks responsible for generating two constellation sequences in a superposition structure.
Specifically, the first JCM block generates the inner constellation sequence $\mathbf{Y}_1=(Y_1^1, Y_1^2, ..., Y_1^n)\in\mathbb{C}^n$, which is intended for both receivers and carries the first level of semantic features to recover the observable source $\mathbf{X}$ and the coarse-grained semantic information $\mathbf{S}_1$.
Here, $n$ denotes the number of channel uses.
The second JCM block generates the outer constellation sequence $\mathbf{Y}_2=(Y_2^1, Y_2^2, ..., Y_2^n)\in\mathbb{C}^n$, which is intended for Receiver 2.
The outer constellation sequence carries the additional semantic features, which together with $\mathbf{Y}_1$, recovers the observable source $\mathbf{X}$ and the fine-grained semantic information $\mathbf{S}_2$.

The first JCM block consists of a basic semantic encoder and a modulator.
The basic semantic encoder $f_{\boldsymbol{\theta}_1}$ with parameters $\boldsymbol{\theta}_1$ extracts and encodes the basic semantic features from $\mathbf{X}$, outputting a $2n$-dimensional real-valued basic EFV $\mathbf{U}_1\in\mathbb{R}^{2n}$
\begin{equation}
    \mathbf{U}_1 = f_{\boldsymbol{\theta}_1}(\mathbf{X}).
\end{equation}
Then, a probabilistic modulator with parameters $\boldsymbol{\rho}_1$ generates $\mathbf{Y}_1$ from $\mathbf{U}_1$ by first learning a transition probability $p(\mathbf{y}_1|\mathbf{u}_1,\boldsymbol{\rho}_1)$ then randomly sampling a sequence according to this transition probability as in our previous work \cite{jcm}.

The second JCM block consists of an enhancement semantic encoder, an LMMSE decorrelator and a modulator.
The enhancement semantic encoder $f_{\boldsymbol{\theta}_2}$ with parameters $\boldsymbol{\theta}_2$ generates an enhanced EFV $\mathbf{U}_2\in\mathbb{R}^{2n}$ from $\mathbf{X}$, which contains the semantic features of $\mathbf{X}$ and $\mathbf{S}_2$
\begin{equation}
    \mathbf{U}_2 = f_{\boldsymbol{\theta}_2}(\mathbf{X}).
\end{equation}
Considering the redundancy in $\mathbf{U}_2$ due to the hierarchical semantics, the LMMSE decorrelator projects $\mathbf{U}_2$ onto the space of $\mathbf{U}_1$, resulting in the successive refinement vector $\mathbf{R}$ of $\mathbf{U}_1$, so that $\mathbf{R}$ and $\mathbf{U}_1$ can be uncorrelated.
A detailed discussion on the LMMSE decorrelator will be further provided in Section III.
Then, in the same manner that $\mathbf{Y}_1$ is generated, a probabilistic modulator with parameters $\boldsymbol{\rho}_2$ learns the transition probability $p(\mathbf{y}_2|\mathbf{r},\boldsymbol{\rho}_2)$ then randomly samples $\mathbf{Y}_2$ based on $\mathbf{R}$.



\begin{figure*}[t]
\centering
\includegraphics[scale=0.7]{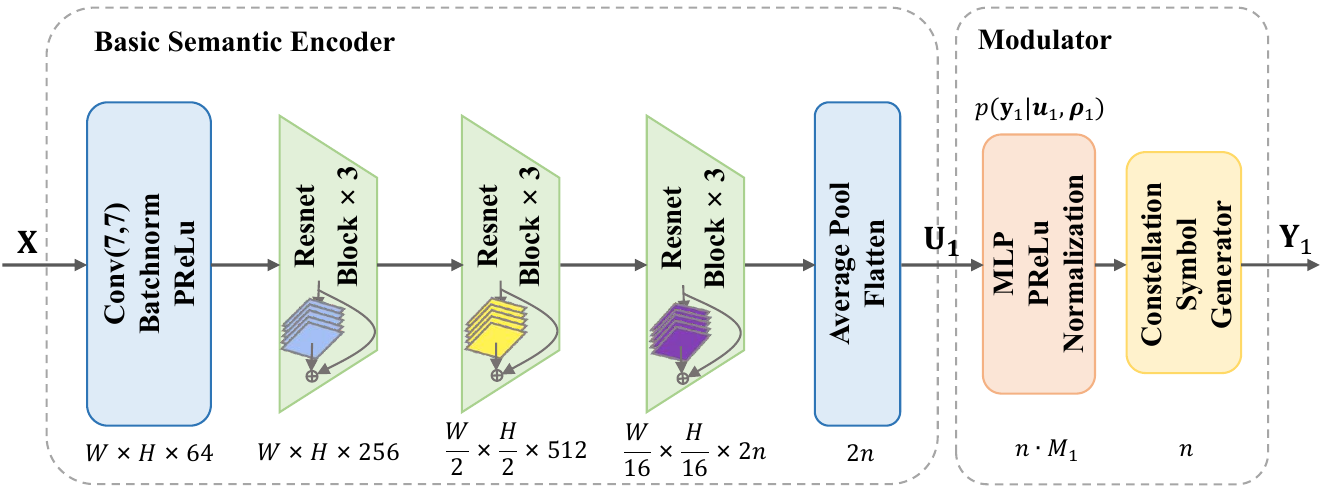}
\caption{NN architecture of the first JCM block.}
\label{nn}
\vspace{-0.3cm}
\end{figure*} 

In this paper, digital transmission is carried out with $M$-QAM modulation by superposing $\mathbf{Y}_1$ and $\mathbf{Y}_2$.
Each element in $\mathbf{Y}_1$ takes values from an $M_1$-QAM constellation $\mathcal{C}_1$, and each element in $\mathbf{Y}_2$ takes values from an $M_2$-QAM constellation $\mathcal{C}_2$, where $M_1$ and $M_2$ satisfy $M_1\times M_2=M$.
The output of the transmitter, which we denote as $\mathbf{Y}\in\mathbb{C}^n$, is formed by
\begin{equation}
    \mathbf{Y}=\sqrt{\alpha}\mathbf{Y}_1+\sqrt{1-\alpha}\mathbf{Y}_2,
    \label{superposition}
\end{equation}
where $\alpha$ denotes the power allocation factor (PAF), and we set $\alpha\in(0.5, 1)$ so that the inner constellation is truly ``inner''.
Note that before superposition, we respectively scale $\mathbf{Y}_1$ and $\mathbf{Y}_2$ so that they meet an average transmit power constraint $P$, \emph{i.e.}, $\frac{\|\mathbf{Y}_1\|^2}{n}=\frac{\|\mathbf{Y}_2\|^2}{n}=P$.
The average transmit power constraint $P$ is again enforced after superposition
\begin{equation}
   \frac{\|\mathbf{Y}\|^2}{n}=P.
\end{equation}
Receiver 1 receives a sequence $\mathbf{ Z}_1=\mathbf{Y}+\boldsymbol{\epsilon}_1$, where $\boldsymbol{\epsilon}_1\sim \mathcal{CN}(\mathbf{0},\sigma^2_1\mathbf{I}_{n\times n})$.
Similarly, Receiver 2 receives a sequence $\mathbf{Z }_2=\mathbf{Y}+\boldsymbol{\epsilon}_2$ with $\boldsymbol{\epsilon}_2\sim \mathcal{CN}(\mathbf{0},\sigma^2_2\mathbf{I}_{n\times n})$ and $\sigma_1 > \sigma_2$.
The channel condition of Receiver $i$ for $i$ in $\{1,2\}$ is characterized by the channel SNR, which is defined as $\frac{P}{\sigma_i^2}$.
Fig.~\ref{examples} displays examples of superposed inner and outer constellations (which we henceforward call \textit{super-constellations}). 
Fig.~\ref{examples}(a), \ref{examples}(b), and \ref{examples}(c) depict the super-constellation resulting from the superposition of two 4QAM constellations, denoted as the 4QAM$\times$4QAM super-constellation. 
Fig.~\ref{examples}(d), \ref{examples}(e), and \ref{examples}(f) illustrate the super-constellation obtained by superposing a 4QAM inner constellation and a 16QAM outer constellation, denoted as the 4QAM$\times$16QAM super-constellation.
As a side note, the $M$-QAM digital modulation can be easily extended to other modulation schemes, such as $M$-PSK.

At each receiver, two neural decoders are deployed in parallel to recover $\mathbf{X}$ and $\mathbf{S}_i$, respectively, from $\mathbf{Z}_i$ ($i \in {1, 2}$).
That is,
\begin{align}
    \mathbf{\hat S}_i&=g_{\boldsymbol{\psi}_i}(\mathbf{Z}_i),\\
    \mathbf{\hat X}_i &= g_{\boldsymbol{\eta}_i}(\mathbf{Z}_i),
\end{align}
where $\boldsymbol{\psi}_i$ parameterizes Receiver $i$'s decoder for classification and $\boldsymbol{\eta}_i$ parameterizes its decoder for image recovery.



The above DeepSCM framework can be easily extended to $K$ ($K>2$) users.
More specifically, consider the observable source $\mathbf{X}$ with implicit hierarchical semantic information $\mathbf{S}_1$, $\mathbf{S}_2$, ..., $\mathbf{S}_K$, forming the Markov chain
\begin{equation}
    \mathbf{S}_1 \to \mathbf{S}_2 \to ... \to \mathbf{S}_K \to \mathbf{X}.
\end{equation}
Receiver $i$, for $i \in [K]$, requires the source data $\mathbf{X}$ and the semantic information $\mathbf{S}_i$.
Similar to the two-user case, there are $K$ JCM blocks at the transmitter.
For each JCM block $i$, for $i \in [K]$, it outputs the feature vector $\mathbf{U}_i\in\mathbb{R}^{2n}$, which contains the semantic features of $\mathbf{X}$, $\mathbf{S}_1$, ..., $\mathbf{S}_i$.
The feature vector $\mathbf{U}_1$ is directly modulated into the constellation sequence  $\mathbf{Y}_1 \in \mathbb{C}^n$, while for $i\ge 2$, $\mathbf{U}_i$ first goes through decorrelation to be projected as the successive refinement vector $\mathbf{R}_i$ of $\mathbf{U}_{i-1}$, which is then modulated to produce $\mathbf{Y}_i\in \mathbb{C}^n$.
Each constellation sequence is normalized before they are superposed to form the output super-constellation sequence $\mathbf{Y}$
\begin{equation}
    \mathbf{Y} = \sum_{i=1}^K \sqrt{\alpha_i} \mathbf{Y}_i,
\end{equation}
where $\alpha_i$ denotes the PAF for $\mathbf{Y}_i$, satisfying $\alpha_1>\alpha_2>...>\alpha_K$ and $\sum_{i=1}^K\alpha_i=1$.
The superposed signal is normalized to the average power of $P$ before transmission.
The receivers are similarly designed to those in the two-user case, with each receiver employing two parallel neural decoders to respectively recover the common source data and the desired-level of semantic information.



\section{Transmitter Design}


The core of our DeepSCM scheme lies in the transmitter design, which aims at extracting uncorrelated semantic features and generating the inner and outer constellation sequences in the superposition structure.
This section is dedicated to describing the components of the transmitter as well as the training algorithm.
We first discuss the basic building block of the transmitter, namely the JCM block.
Then, in Section III-B, we explain the technical detail of the aforementioned LMMSE decorrelator.
Finally, in Section III-C, we present the training algorithm of the whole system.

\subsection{Joint Coding-Modulation}

Extending our previous work\cite{jcm,bo2023joint} to a broadcast setting, we use two JCM blocks to enable the digital modulation of the two EFVs.
The NN architecture of the first JCM block is shown in Fig.~\ref{nn}.
The basic semantic encoder includes multiple Resnet blocks\cite{resnet} to map an image $\mathbf{X}$ of dimension $W\times H\times C$ into the basic EFV $\mathbf{U}_1$ of dimension $2n$, where $W$, $H$ and $C$ respectively denote the width, height and channel of the image, and we have $d=W\times H\times C$.

We then modulate $\mathbf{U}_1$ to obtain $\mathbf{Y}_1$.
The modulation process is learned as a probabilistic model to avoid the inherent nondifferentiability problem of the digital modulation.
A multi-layer perceptron (MLP) with a $\text{PReLu}(\cdot)$ activation function and a normalization layer outputs the transition probability, denoted as $p(\mathbf{y}_1|\mathbf{u}_1,\boldsymbol{\rho}_1)$.
Since $\mathbf{Y}_1=(Y_1^1, Y_1^2, ..., Y_1^n)$ is discretely distributed with $M_1^n$ possible values, the transition probability is a discrete probability distribution with $M_1^n$ categories.
To simplify the learning process, we model each element of $\mathbf{Y}_1$ to be conditionally independent, reducing the total number of probability categories to be learned to $n\cdot M_1$.
That is, for each element of $\mathbf{Y}_1$, the MLP respectively outputs $M_1$ probabilities.
Based on this probability distribution, the constellation symbol generator samples a constellation symbol for this element by using the Gumbel-Softmax method\cite{jang2016softgumbel} as detailed below.

As a differentiable sampling technique, the Gumbel-Softmax method is used to reparameterize the discrete variable $\mathbf{Y}_1$ as a deterministic function of an independent random variable and the transition probability, thereby enabling the backpropagation through one sample of $\mathbf{Y}_1$. 
Specifically, in the forward pass, the sampling process of $Y_1^j$, where $j\in[n]$, from the transition probability $p(\mathbf{y}_1|\mathbf{u}_1,\boldsymbol{\rho}_1)$ can be expressed as
\begin{equation}
    y_1^j=\mathbf{c}^T\cdot \text{one-hot}\left (\underset{m\in[M_1]}{\mathrm{argmax}}[\tau_{jm}+\log q_{jm|\mathbf{u}_1,\boldsymbol{\rho}_1}]\right ),
\end{equation}
where $y_1^j$ denotes a sample of $Y_1^j$, $\mathbf{c}=(c_1, c_2, ..., c_{M_1})$ is the vector composed of all the constellation symbols of the $M_1$-QAM, $\tau_{jm}$ is an independent random variable with Gumbel distribution $\tau_{jm} \sim \mathrm{Gumbel}(0,1)$, and $q_{jm|\mathbf{u}_1,\boldsymbol{\rho}_1}$ is defined as 
$q_{jm|\mathbf{u}_1,\boldsymbol{\rho}_1}=P(Y_1^j=c_m|\mathbf{u}_1,\boldsymbol{\rho}_1)$.
The function $\text{one-hot}(x)$ converts the index $x$ into its one-hot form, in which the $x$-th element is equal to 1 and the rest are equal to 0.
The dot product of $\mathbf{c}^T$ and the one-hot vector samples the constellation symbol for $Y_1^j$.
In the backpropagation, the non-differentiable $\mathrm{argmax}$ function is approximated by the differentiable $\mathrm{softmax}$ function, enabling the stochastic gradient descent (SGD) algorithm. 
Note that the above sampling process is proven to obtain constellation sequences whose empirical distribution matches their probability distribution\cite{bo2023joint}.

The second JCM block follows the same architecture as the first one, except that the enhanced EFV $\mathbf{U}_2$ should be decorrelated with $\mathbf{U}_1$ before modulation.

\subsection{LMMSE Decorrelator}

Due to the hierarchical relationship of the semantics, the enhanced EFV $\mathbf{U}_2$ is highly correlated with the basic EFV $\mathbf{U}_1$.
Directly superposing them will cause redundancy in transmission and decrease the transmission efficiency.

To achieve maximum transmission efficiency, it is important that the EFVs being superposed are independent.
Thus, our goal is to remove from $\mathbf{U}_2$ the information that is dependent on $\mathbf{U}_1$ so that we can obtain a reduced-rate EFV that carries supplementary information from $\mathbf{U}_1$.
We call this reduced-rate EFV the successive refinement vector of $\mathbf{U}_1$, denoted as $\mathbf{R}$.
In other words, we want to find both a function $F(\cdot)$ and a vector $\mathbf{R}$, so that $\mathbf{U}_2$ can be written as
\begin{equation}
    \mathbf{U}_2 = F(\mathbf{U}_1)+\mathbf{R},
\end{equation}
where $\mathbf{R}$ should have smaller entropy rate since it carries less information.
For simplicity, in this paper we choose a linear model for $F(\cdot)$. 
Hence, $\mathbf{U}_2$ can be written as
\begin{align}
    \mathbf{U}_2
    &=\mathbf{WU}_1+\mathbf{b}+\mathbf{R},
    \label{linear}
\end{align}
where $\mathbf{W}\in\mathbb{R}^{2n\times 2n}$ and $\mathbf{b}\in\mathbb{R}^{2n}$ are parameters to be determined so that a reduced-rate $\mathbf{R}$ can be obtained.
However, it is hard to directly minimize the entropy of $\mathbf{R}$, given the challenge of estimating the entropy rate of a random variable.
The following proposition presents an alternative approach to finding a low-rate $\mathbf{R}$. 
\begin{proposition}
\label{upperbound}
The entropy of $\mathbf{R}$ is upperbounded by
\begin{equation}
h(\mathbf{R})\le n\log \frac{\pi e}{n} \mathbf{E}[||\mathbf{R}||_2^2],
\end{equation}
where the equality holds if each element of $\mathbf{R}$ follows an i.i.d. Gaussian distribution with zero mean.
\end{proposition}

The proof can be found in Appendix A. As can be seen, the upperbound of the entropy of $\mathbf{R}$ is decided by $\mathbf{E}[||\mathbf{R}||_2^2]$, \emph{i.e.}, the expected $l_2$-norm of $\mathbf{R}$.
Therefore, minimizing the entropy of $\mathbf{R}$ can be translated into minimizing $\mathbf{E}[||\mathbf{R}||_2^2]$, which is much more tractable. 
By using the linear model in \eqref{linear}, the term $\mathbf{E}[||\mathbf{R}||_2^2]$ can be expanded as
\begin{equation}
    \mathbf{E}[||\mathbf{R}||_2^2]=\mathbf{E}[(\mathbf{U}_2-\mathbf{WU}_1-\mathbf{b})^T(\mathbf{U}_2-\mathbf{WU}_1-\mathbf{b})].
\end{equation}
The optimization problem is thus established as
\begin{equation}
    \underset{\mathbf{W},\mathbf{b}}{\mathrm{min}} \ \ \mathbf{E}[(\mathbf{U}_2-\mathbf{W}\mathbf{U}_1-\mathbf{b})^T(\mathbf{U}_2-\mathbf{W}\mathbf{U}_1-\mathbf{b})].
\label{reg}
\end{equation}

Notice that, problem \eqref{reg} is essentially an LMMSE estimation problem of $\mathbf{U}_2$ given $\mathbf{U}_1$, whose global optimal solution is given by
\begin{align}
    \mathbf{W}^\ast&=\textit{Cov}[\mathbf{U}_2,\mathbf{U}_1]\textit{Var}[\mathbf{U}_1]^{-1},\label{w}\\
    \mathbf{b}^\ast&=\mathbf{E}[\mathbf{U}_2]-\mathbf{W}^\ast \mathbf{E}[\mathbf{U}_1].
\end{align}
Following the orthogonality principle of linear estimators \cite[p.386]{kay1993fundamentals}, the global optimal $\mathbf{R}^\ast$, which serves as the estimation error, is linearly uncorrelated with $\mathbf{U}_1$, \emph{i.e.}, 
\begin{equation}
    \mathbf{E}[\mathbf{U}_1\mathbf{R}^{\ast T}]=\mathbf{0}.
    \label{uncor}
\end{equation}

The direct use of LMMSE estimation 
is computationally demanding since it requires the computation of the inversion of large matrices. 
As such, we utilize a one-layer MLP for $F(\cdot)$ with input $\mathbf{U}_1$ and output
${\mathbf{W}}\mathbf{U}_1+\mathbf{b}$, where $\mathbf{W}$ and $\mathbf{b}$ respectively represent the weights and bias of the MLP.
The MLP-based LMMSE decorrelator is both computationally efficient and compatible with the mini-batch SGD algorithm.

\begin{algorithm}
   \caption{\small DeepSCM training algorithm.}\label{algorithm}
    \small

\textbf{Stage 1, input:}
Image samples $\mathbf{X}$ at Tx, image samples $\mathbf{X}$ and coarse classification labels $\mathbf{S}_1$ at Rx1.

\While{stop criterion is not met}{
\textbf{Forward pass}\\
Tx computes and broadcasts $\mathbf{Y}$;\\
Rx1 receives $\mathbf{Z}_1$, and computes $g_{\boldsymbol{\eta}_1}(\mathbf{Z}_1)\to \mathbf{\hat X}_1$, $g_{\boldsymbol{\psi}_1}(\mathbf{Z}_1)\to \mathbf{\hat S}_1$, and the loss $\mathcal{L}_1$ by \eqref{loss 1};\\
\textbf{Backward pass}\\
Rx1 trains $\boldsymbol{\psi}_1$,  $\boldsymbol{\eta}_1$, then transmits the gradients of the first layer of $g_{\boldsymbol{\eta}_1}$ and $g_{\boldsymbol{\psi}_1}$ through an error-free link to Tx;\\
Tx trains $\boldsymbol{\theta}_1$,  $\boldsymbol{\rho}_1$;}

\textbf{Stage 2, input:} Image samples $\mathbf{X}$ at Tx, image samples $\mathbf{X}$ and fine classification labels $\mathbf{S}_2$ at Rx2.

\While{stop criterion is not met}{
\textbf{Forward pass}\\
Tx computes and broadcasts $\mathbf{Y}$;\\
Rx2 receives $\mathbf{Z}_2$, and computes $g_{\boldsymbol{\eta}_2}(\mathbf{Z}_2)\to \mathbf{\hat X}_2$, $g_{\boldsymbol{\psi}_2}(\mathbf{Z}_2)\to \mathbf{\hat S}_2$, and the loss $\mathcal{L}_2$ by \eqref{loss 2};\\
\textbf{Backward pass}\\
Rx2 trains $\boldsymbol{\psi}_2$,  $\boldsymbol{\eta}_2$, and transmits the gradients of the first layer of $g_{\boldsymbol{\eta}_2}$ and $g_{\boldsymbol{\psi}_2}$ through an error-free link to Tx;\\
Tx trains $\boldsymbol{\theta}_2$,  $\boldsymbol{\rho}_2$;\\}

\textbf{Stage 3, input:} Image samples $\mathbf{X}$ at Tx, image samples $\mathbf{X}$ and coarse classification labels $\mathbf{S}_1$ at Rx1, image samples $\mathbf{X}$ and fine classification labels $\mathbf{S}_2$ at Rx2.

\While{stop criterion is not met}{
\textbf{Forward pass}\\
Tx computes and broadcasts $\mathbf{Y}$;\\
Rx1 receives $\mathbf{Z}_1$, and computes $g_{\boldsymbol{\eta}_1}(\mathbf{Z}_1)\to \mathbf{\hat X}_1$, $g_{\boldsymbol{\psi}_1}(\mathbf{Z}_1)\to \mathbf{\hat S}_1$, and the loss $\mathcal{L}_1$ by \eqref{loss 1};\\
Rx2 receives $\mathbf{Z}_2$, and computes $g_{\boldsymbol{\eta}_2}(\mathbf{Z}_2)\to \mathbf{\hat X}_2$, $g_{\boldsymbol{\psi}_2}(\mathbf{Z}_2)\to \mathbf{\hat S}_2$, and the loss $\mathcal{L}_2$ by \eqref{loss 2};\\
\textbf{Backward pass}\\
Rx1 trains $\boldsymbol{\psi}_1$,  $\boldsymbol{\eta}_1$, and transmits the gradients of the first layer of $g_{\boldsymbol{\eta}_1}$ and $g_{\boldsymbol{\psi}_1}$ through an error-free link to Tx;\\
Rx2 trains $\boldsymbol{\psi}_2$,  $\boldsymbol{\eta}_2$, and transmits the gradients of the first layer of $g_{\boldsymbol{\eta}_2}$ and $g_{\boldsymbol{\psi}_2}$ through an error-free link to Tx;\\
Tx trains $\boldsymbol{\theta}_2$,  $\boldsymbol{\rho}_2$ using gradient descent with $\mathcal{L}_3$ in \eqref{loss 3}, \emph{i.e.}, $\mathcal{L}_1+\beta\mathcal{L}_2$;}
\textbf{Output:} trained parameters $\boldsymbol{\theta}_1$, $\boldsymbol{\rho}_1$, $\boldsymbol{\psi}_1$, $\boldsymbol{\eta}_1$, $\boldsymbol{\theta}_2$, $\boldsymbol{\rho}_2$, $\mathbf{W}$, $\mathbf{b}$, $\boldsymbol{\psi}_2$, $\boldsymbol{\eta}_2$.
\end{algorithm}

\subsection{Training Strategy}

As can be seen from Fig.~\ref{overview}, there are a total of nine NN modules used in the proposed DeepSCM scheme for different purposes. 
The intricate design of this framework makes it difficult to ensure the convergence of the network training. 
Therefore, we adopt a three-stage training strategy.
In the first stage, the encoding and decoding NNs for the generation of the inner constellation sequence are trained, namely the first JCM block and the decoders at Receiver 1.
Then in the second stage, we further train the encoding and decoding NNs for the generation of the outer constellation sequence, including the second JCM block and the decoders at Receiver 2 while freezing other parameters.
In the last training stage, the whole system is fine-tuned jointly to further improve its overall performance.


Specifically, the loss function for the first training stage is defined as
\begin{equation} \mathcal{L}_1(\boldsymbol{\theta}_1,\boldsymbol{\rho}_1,\boldsymbol{\psi}_1,\boldsymbol{\eta}_1)= D_s(\mathbf{S}_1,\mathbf{\hat S}_1)+\lambda_1\cdot D_x(\mathbf{X},\mathbf{\hat X}_1),
\label{loss 1}
\end{equation}
where $D_s(\cdot, \cdot)$ denotes the distortion measure for semantic information recovery, $D_x(\cdot, \cdot)$ denotes the distortion measure for image recovery, and $\lambda_1$ represents the hyperparameter to balance the two distortions, as used in many previous works\cite{image1, semantic:image3}.
In our work, we employ the commonly used cross entropy (CE) \cite{jcm, gunduz2019jscc1} as the distortion measure for classification, \emph{i.e.}, $D_s(\mathbf{S}_1,\mathbf{\hat S}_1)=\text{CE}(\mathbf{S}_1,\mathbf{\hat S}_1)$. 
For image recovery, two distortion measures are considered.
One is mean square error (MSE), which is used when the image recovery performance is evaluated by peak signal-to-noise ratio (PSNR). 
When the multi-scale structural similarity index (MS-SSIM) is used as the performance metric, the distortion measure is set to be $1-$ MS-SSIM.

Similarly, the loss function of the second stage can be defined as
\begin{align}
    &\ \ \ \ \ \ \ \ \  \mathcal{L}_2(\boldsymbol{\theta}_2,\boldsymbol{\rho}_2,\mathbf{W},\mathbf{b},\boldsymbol{\psi}_2,\boldsymbol{\eta}_2)=\nonumber\\
&D_s(\mathbf{S}_{2},\mathbf{\hat{S}}_{2})+\lambda_2\cdot D_x(\mathbf{X},\mathbf{\hat{X}}_2)+\lambda_3\cdot\|\mathbf{R}\|^2_2,
\label{loss 2}
\end{align}
where the objective in \eqref{reg}, \emph{i.e.}, the $l_2$-norm of the successive refinement vector $\mathbf{R}$, is included as an additional regularizer, and $\lambda_2$ and $\lambda_3$ are two hyperparameters.
Finally, the loss function in the fine-tuning stage is the combination of $\mathcal{L}_1$ and $\mathcal{L}_2$, expressed as
\begin{equation}
\mathcal{L}_3(\boldsymbol{\theta}_1,\boldsymbol{\rho}_1,\boldsymbol{\psi}_1,\boldsymbol{\eta}_1,\boldsymbol{\theta}_2,\boldsymbol{\rho}_2,\mathbf{W},\mathbf{b},\boldsymbol{\psi}_2,\boldsymbol{\eta}_2)=\mathcal{L}_1 + \beta \cdot \mathcal{L}_2,
\label{loss 3}
\end{equation}
where $\beta$ is used to balance the importance of the two receivers.

The overall training procedure is outlined in Algorithm \ref{algorithm}.
More specifically, during each training iteration, the transmitter randomly selects a source sample from the dataset, processes it through the NN, and broadcasts the output sequence to the receivers. 
The environmental condition is sampled to enable the training process.
Each receiver then reconstructs the source sample and the semantic information, computes the loss, and backpropagates the gradients to update NN parameters using the SGD algorithm. 
During backpropagation, each receiver sends the gradients of its first layer through an error-free link to the transmitter, so that the transmitter can compute the gradients of its own NN parameters.

\section{Experiment Results}

In this section, we validate the advantages of DeepSCM via experimental simulation.
We first present the experiment settings in Section IV-A.
Then, we compare the performance of DeepSCM with benchmarks across various transmission rates and channel disparity levels respectively in Section IV-B and IV-C. 
The robustness of DeepSCM against variations in channel conditions is further demonstrated in Section IV-D.
Additionally, we conduct experiments in Section IV-E examining the impact of PAF in the super-constellation.
Finally, we provide discussion on the computation complexity of the proposed method in Section IV-F.

\begin{table*}[t]
    \centering
    \renewcommand{\arraystretch}{1.1}
    \caption{NN architecture of the proposed scheme, where $n$ is the number of channel uses.}
    \begin{tabular}{c|c|c}
    \hline
      & \textbf{Layer} & \textbf{Output Dimension} \\
    \hline
    \multirow{5}{*}{\makecell[c]{\textbf{Basic Semantic Encoder}$\ \setminus$\\
    \textbf{Enhancement Semantic Encoder}}} & Conv + BatchNorm + PReLU & 32$\times$32$\times$64\\
    \multirow{5}{*}{} & Resnet Block $\times$ 3 & 32$\times$32$\times$256\\
    \multirow{5}{*}{}& Resnet Block $\times$ 3 & 16$\times$16$\times$512\\
   \multirow{5}{*}{}& Resnet Block $\times$ 3 & 2$\times$2$\times$2$n$\\
    \multirow{5}{*}{}& Average Pool + Flatten & 2$n$\\
    \hline
    \textbf{Modulator 1} & MLP + PReLU & $n\cdot 2\sqrt{M_1}$ \\
    \hline
    \textbf{LMMSE Decorrelator} & MLP  & $2n$ \\
    \hline
    \textbf{Modulator 2} & MLP + PReLU & $n\cdot 2\sqrt{M_2}$ \\
    \hline
    \textbf{Semantic Decoder for Classification at Rx1} & Spinalnet Block $\times$ 4 + MLP & 20 \\
    \hline
    \textbf{Semantic Decoder for Classification at Rx2} & Spinalnet Block $\times$ 4 + MLP & 100 \\
    \hline
    \multirow{5}{*}{\makecell[c]{\textbf{Semantic Decoder for}\\ \textbf{Image Recovery}\\ \textbf{at Rx1 $\setminus$ Rx2}}} & Conv + BatchNorm + ReLU & 4$\times$4$\times$512 \\
    \multirow{5}{*}{} & Resnet Block $\times$ 2 & 4$\times$4 $\times$512\\
    \multirow{5}{*}{} & Reshape + Conv + BatchNorm + ReLU & 16$\times$16$\times$256 \\
    \multirow{5}{*}{} & Resnet Block $\times$ 2 & 16$\times$16$\times$256\\
    \multirow{5}{*}{} & Reshape + Conv + BatchNorm + Sigmoid & 32$\times$32$\times$3\\
    \hline
    \end{tabular}
    \label{architecture}
    \vspace{-0.3cm}
\end{table*}

\begin{table}[t]
\renewcommand{\arraystretch}{1.1}
    \centering
    \caption{Learning rate and training epochs for each stage.}
    \begin{tabular}{c|c|c|c}
    \hline
         \textbf{Stage}& \textbf{Training Epochs} & \textbf{Initial LR} & \textbf{Scheduler}\\
         \hline
        \textbf{1} & 100 & 2e-4 & \multirow{3}{*}{\makecell[c]{Cosine Annealing\\
    Warm Restarts\\with Final LR=1e-5}}\\
        \textbf{2} & 200 & 2e-4 \\
        \textbf{3} & 100 & 5e-5 \\
        \hline
    \end{tabular}
    \label{lr}
    \vspace{-0.3cm}
\end{table}

\subsection{Experiment Settings}

\subsubsection{Dataset and Channel}

Our experiments are conducted on the CIFAR100 dataset\cite{krizhevsky2009cifar}, which includes
50000 training images and 10000 test images. 
The resolution of each image is $32\times32$. 
All images are classified into 20 super-categories, and the images in each super-category are further classified to 5 sub-categories, resulting a total of 100 categories \cite{krizhevsky2009cifar}. 
The 20-category classification label of each image stands for its coarse-grained semantic information $\mathbf{S}_1$ while the 100-category classification label of each image stands for its fine-grained semantic information $\mathbf{S}_{2}$.
Throughout the simulation, we consider two-user Gaussian broadcast channels. 
If not specified otherwise, the SNR of receiver 1 is fixed at -5 dB while the SNR of receiver 2 varies between 0 dB and 20 dB.  
Receiver 1 intends to recover the original image as well as to perform the 20-category coarse classification while Receiver 2 intends to recover the original image and to perform the 100-category fine classification.

\subsubsection{NN Architecture and Hyperparameters}
The transmitter design has been presented in Section III.
The semantic decoders for image recovery at both receivers have the same NN architecture, which consists of Resnet blocks combined with the depth-to-space operation to perform the upsampling.
Spinal-net\cite{spinal} is adopted for both coarse and fine classification.
Table~\ref{architecture} presents the details of the NN architecture we use in this paper.
We employ the Adam optimizer for the training.
The learning rate (LR) schedule is shown in Table~\ref{lr}.
The experiments are conducted using two super-constellations: 4QAM$\times$4QAM and 4QAM$\times$16QAM.
We set the value of the PAF $\alpha$ such that the super-constellations form rectangular QAM constellations.
Specifically, for 4QAM$\times$4QAM, we set $\alpha=0.80$, and for 4QAM$\times$16QAM, we set $\alpha=0.76$.
All the experiments are performed on Intel Xeon Silver 4214R CPU, and 24 GB Nvidia GeForce RTX 3090 Ti graphics card with Pytorch powered with CUDA 11.4.

\subsubsection{Benchmarks}

We compare the proposed DeepSCM with the conventional unstructured coded modulation (CM) scheme, where only one semantic encoder with rectangular M-QAM modulation is utilized at the transmitter.
In the CM scheme, the NN architectures of the semantic encoder, modulator, and the two receivers are the same as their counterparts in DeepSCM.
Three different training methods shall be used for the CM scheme as three different benchmarks.

\begin{itemize}

\item \emph{CM Trained Jointly}: 
In this scheme, the transmitter and the two receivers are jointly trained, where the loss function is set to be a weighted sum of the distortion measures of both receivers.
This training scheme tries to balance the performance of the two receivers.

\item \emph{CM Trained with Rx1 (Performance Upperbound for Rx1)}: 
Following the training method in \cite{one2many}, in this scheme, the transmitter and Receiver 1 are first jointly trained in the absence of Receiver 2.
After the transmitter is trained and fixed, Receiver 2 is then trained alone to achieve its best possible performance.
Note that this scheme serves as the performance upperbound for Receiver 1 under the CM scheme.
%

\item \emph{CM Trained with Rx2 (Performance Upperbound for Rx2)}:
This training scheme follows the same approach as ``CM trained with Rx1'' but exchanges the role of Receiver 1 and Receiver 2.
Similarly, it serves as the performance upperbound for Receiver 2 under the CM scheme.

\end{itemize}

Additionally, we compare our DeepSCM scheme with classical separate source and channel coding (SSCC) schemes.
Three SSCC schemes are considered as baselines, namely: JPEG2000 with capacity-achieving channel code (abbreviated as ``JPEG2000+Capacity''), Better Portable Graphics (BPG) with capacity-achieving channel code (``BPG+Capacity''), and a more practical scheme, BPG with low density parity check (LDPC) codes (``BPG+LDPC'').
The LDPC codes are from the widely-used IEEE 802.11n standard, with block length 648 bits for rate $\frac{1}{2}$,  $\frac{2}{3}$ and $\frac{3}{4}$.
As a two-user broadcast channel is considered in this paper, the SSCC baselines are implemented with two multiple-access strategies. 
One is the single-user (SU) strategy where the coding is designed for one user only, and the other is time division multiple access (TDMA), where each user occupies half of the channel uses and the coding is specifically designed for that user during its allocated channel uses. 



\begin{figure*}[t]
     \centering
     \begin{subfigure}[b]{0.98\textwidth}
         \centering
         \includegraphics[width=0.95\textwidth]{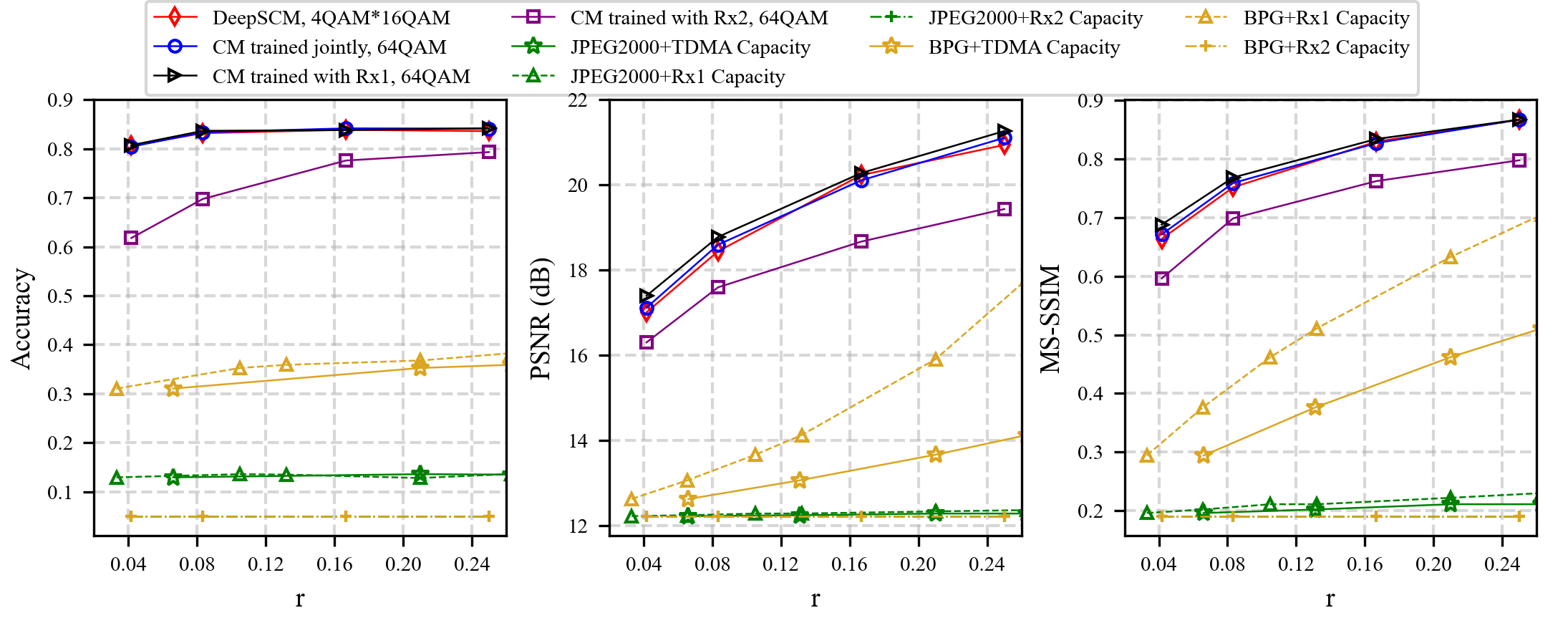}
         \caption{Performance comparison for Rx1: (left) classification accuracy; (center) PSNR; (right) MS-SSIM.}
         \label{rate_psnr_64}
     \end{subfigure}
     \begin{subfigure}[b]{0.98\textwidth}
         \centering
         \includegraphics[width=0.95\textwidth]{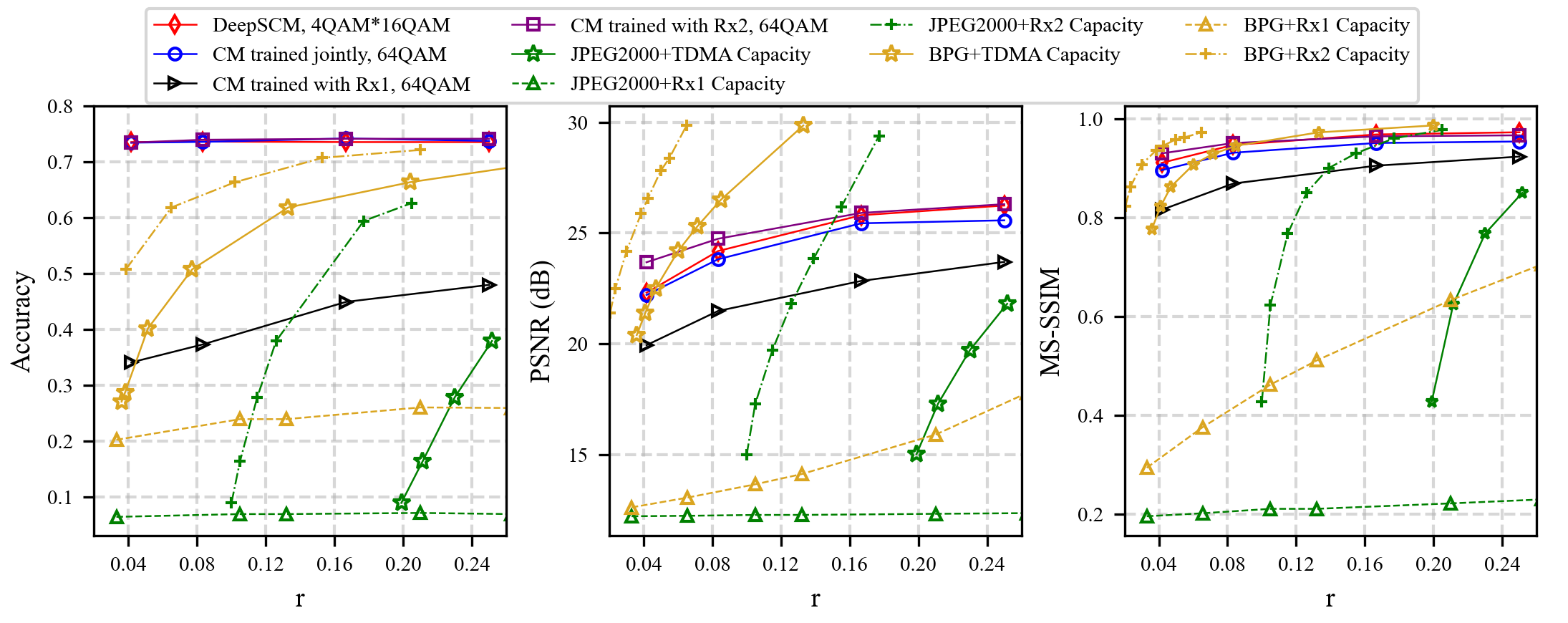}
         \caption{Performance comparison for Rx2: (left) classification accuracy; (center) PSNR; (right) MS-SSIM.}
         \label{rate_acc_64}
     \end{subfigure}
    \caption{Performance of different schemes at varying transmission rates, with 4QAM$\times$16QAM super-constellation.}
    \label{figure:rate 64}
    \vspace{-0.4cm}
\end{figure*}

\subsubsection{Performance Metrics}
For the coarse and fine classification tasks, we use classification accuracy to evaluate the performance.
For the image recovery task, we utilize the pixel-wise performance metric PSNR and the perceptual metric MS-SSIM\cite{wang2003multiscale} to evaluate the performance.

We define the transmission rate $r$ as the ratio between the number of channel uses and the dimension of the images, \emph{i.e.},
\begin{equation}
    r = \frac{n}{d},
\end{equation}
where for CIFAR100 one has $d=32\times32\times3$.

\begin{figure*}[t]
     \centering
     \begin{subfigure}[b]{0.98\textwidth}
         \centering
         \includegraphics[width=0.95\textwidth]{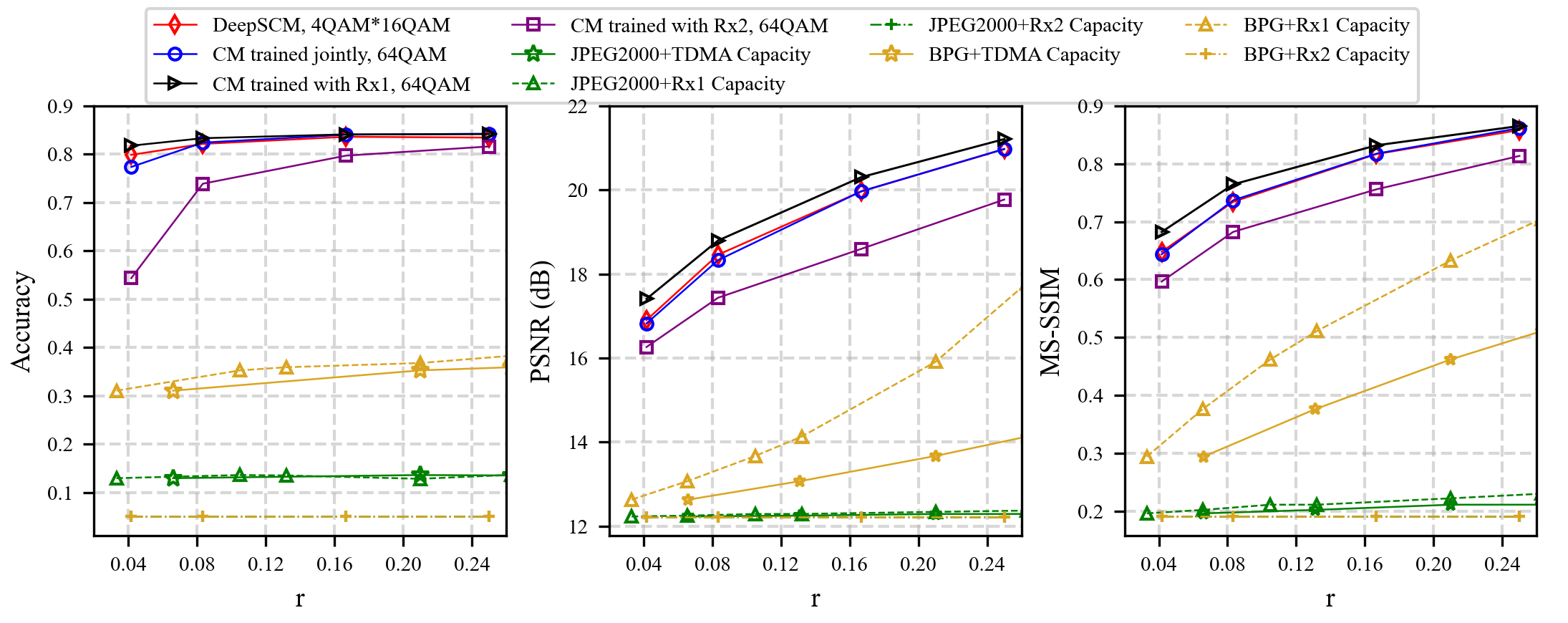}
         \caption{Performance comparison for Rx1: (left) classification accuracy; (center) PSNR; (right) SSIM.}
         \label{rate_psnr_16}
     \end{subfigure}
     \begin{subfigure}[b]{0.98\textwidth}
         \centering
         \includegraphics[width=0.95\textwidth]{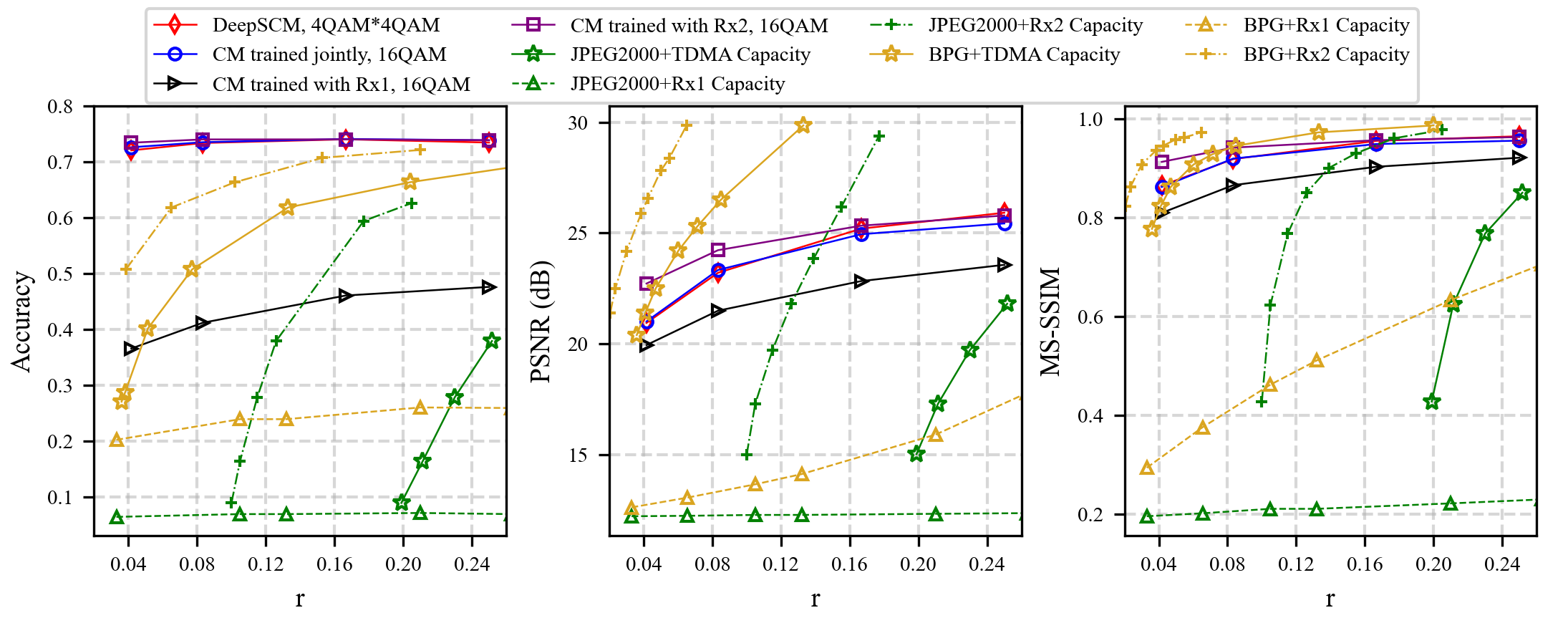}
         \caption{Performance comparison for Rx2: (left) classification accuracy; (center) PSNR; (right) SSIM.}
         \label{rate_acc_16}
     \end{subfigure}
    \caption{Performance of different schemes at varying transmission rates, with 4QAM$\times$4QAM super-constellation.}
    \label{figure:rate 16}
    \vspace{-0.4cm}
\end{figure*}

\begin{figure*}[t]
\centering
\includegraphics[scale=0.53]{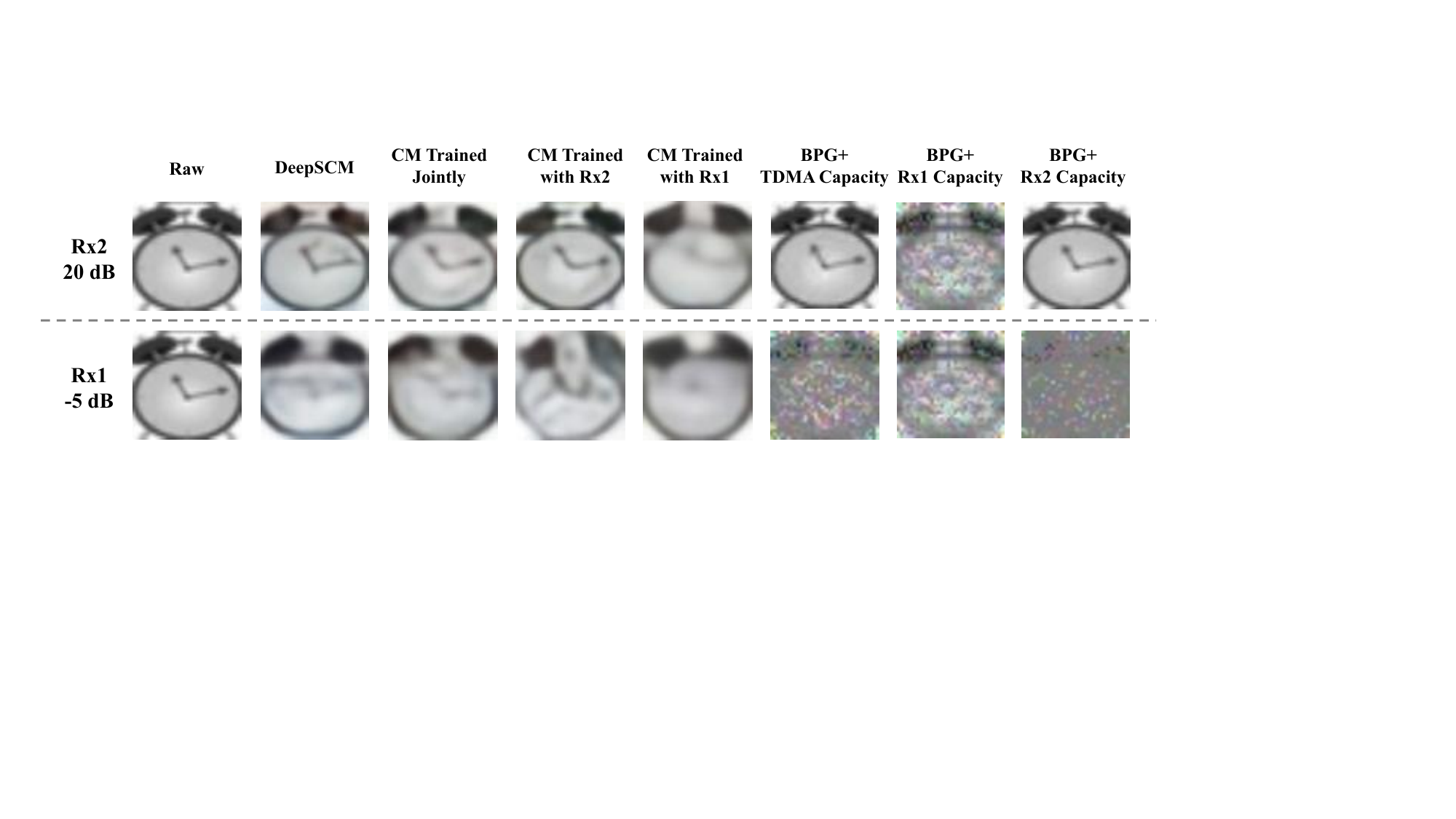}
\caption{Visual examples of the image recovered by different schemes for Rx1 and Rx2 with $M=64$ and $r=0.167$.}
\label{example}
\vspace{-0.4cm}
\end{figure*} 

\begin{figure*}[t]
     \centering
     \begin{subfigure}[b]{0.95\textwidth}
         \centering
         \includegraphics[width=0.95\textwidth]{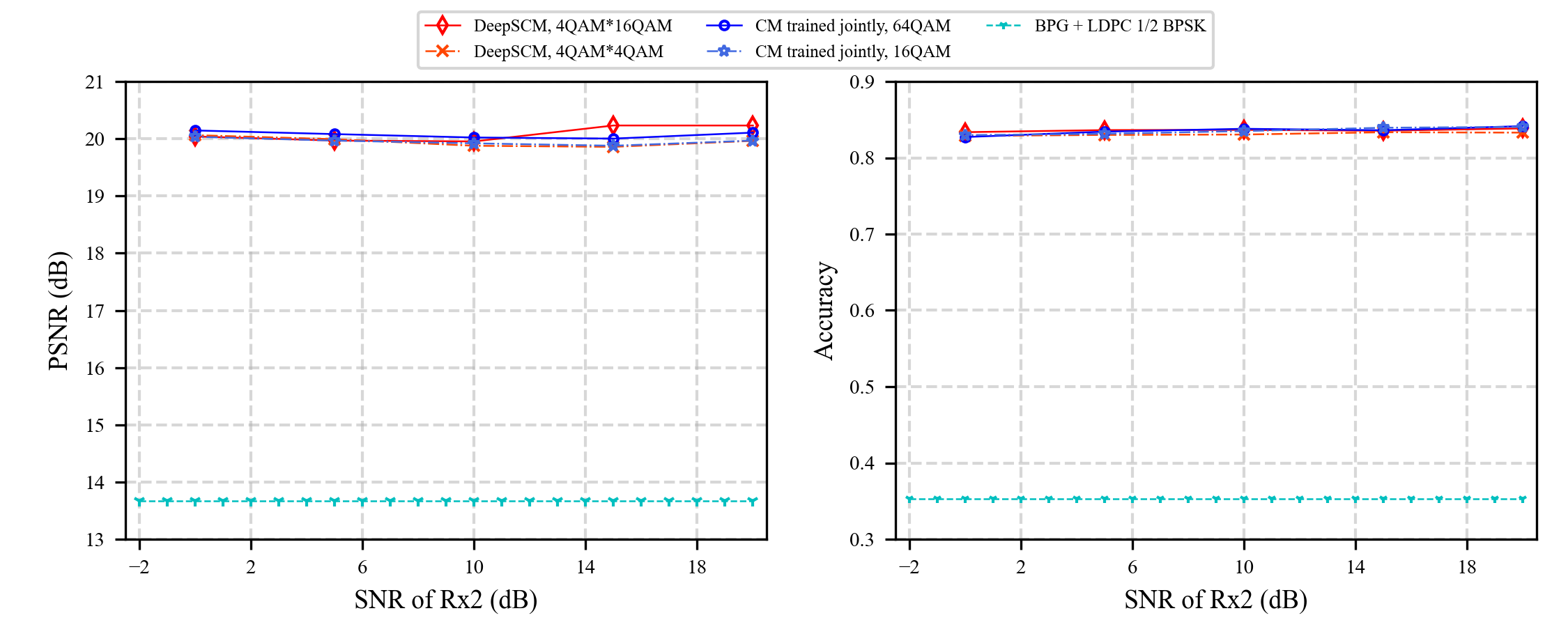}
         \caption{Performance comparison for Rx1: (left) PSNR; (right) classification accuracy.}
         \label{snr_psnr}
     \end{subfigure}
     \begin{subfigure}[b]{0.95\textwidth}
         \centering
         \includegraphics[width=0.95\textwidth]{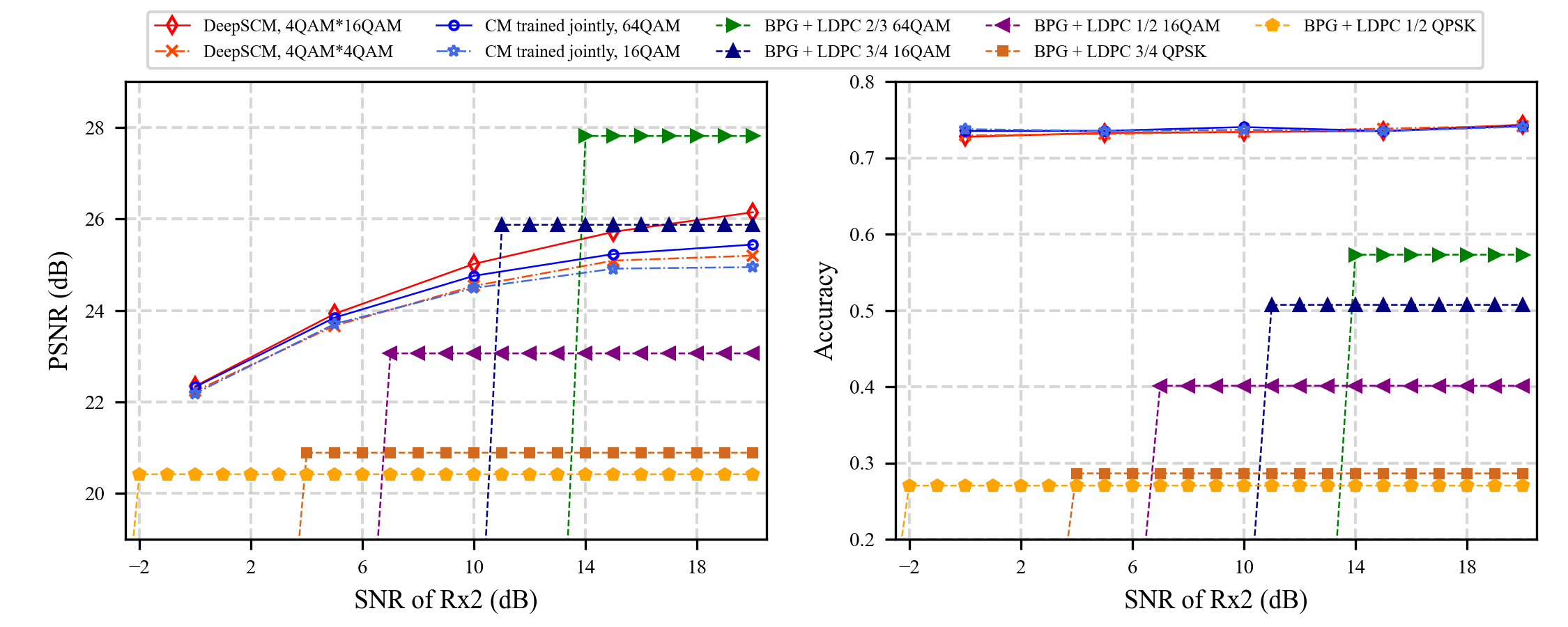}
         \caption{Performance comparison for Rx2: (left) PSNR; (right) classification accuracy.}
         \label{snr_acc}
     \end{subfigure}
    \caption{Performance comparison at different channel SNRs of Rx2. The SNR of Rx1 is constantly set as -5 dB.}
    \label{figure:snr}
\vspace{-0.4cm}
\end{figure*}

\subsection{Performances at Varying Transmission Rates}


In this subsection, we compare the performance of DeepSCM and the benchmarks when the transmission rate varies from 0.04 (128 channel uses) to 0.25 (768 channel uses).
If not specified otherwise, we set the channel SNR of Receiver 1 as $-5$ dB and that of Receiver 2 as 20 dB.

Fig.~\ref{figure:rate 64} focuses on the case when a 4QAM$\times$16QAM super-constellation is used for our proposed DeepSCM scheme. 
For fair comparison, all three CM schemes adopt 64QAM modulation. 
Fig.\ref{figure:rate 64}(a) and Fig.\ref{figure:rate 64}(b) illustrate the performance comparison for Receiver 1 and Receiver 2, respectively, in terms of classification accuracy, PSNR, and MS-SSIM.
We first compare the proposed DeepSCM with the three different CM benchmarks.
Overall, DeepSCM's classification and image recovery performance for both receivers is close to their respective upperbounds.
For classification accuracy, the proposed scheme and the ``CM trained jointly'' scheme exhibit similar performance on both receivers, both approaching the CM upperbound (achieved by ``CM trained with Rx1'' on Receiver 1 and by ``CM trained with Rx2'' on Receiver 2).
Note that while being the performance upperbound for Receiver 1, the ``CM trained with Rx1'' scheme performs poorly on receiver 2, and and vice versa for ``CM trained with Rx2''.
Notably, the classification performances of the two receivers are not directly comparable as Receiver 1 only classifies 20 categories while Receiver 2 classifies 100.
For PSNR, DeepSCM approaches the upperbound of ``CM trained with Rx1'', performs comparably to the ``CM trained jointly'' scheme, and outperforms ``CM trained with Rx2'' by more than 1 dB for Receiver 1.
For Receiver 2, DeepSCM's PSNR performance approaches the upperbound (``CM trained with Rx2''), especially at higher rates, outperforming the ``CM trained jointly'' scheme by 0 to 0.6 dB and the ``CM trained with Rx1'' scheme by more than 2 dB.
The MS-SSIM curves follow the same trends as the PSNR curves, with DeepSCM simultaneously approaching the performance upperbounds of both receivers and outperforming the ``CM trained jointly'' scheme.
For example, at $r=0.25$, DeepSCM achieves an MS-SSIM of 0.972 for Receiver 2, while the ``CM trained jointly'' scheme achievs an MS-SSIM of 0.954, 1.85$\%$ lower than DeepSCM.

Next, we compare the performance of the proposed DeepSCM with the classic ``BPG/JPEG2000+Capacity''  baselines using both SU coding strategy and TDMA strategy.
It can be generally observed that the proposed scheme better balances the performances of the two receivers. 
In the SU coding scheme, optimizing coding for one receiver results in suboptimal performance for the other. 
For example, the ``BPG+Rx2 capacity'' scheme outperforms the proposed scheme in PSNR and MS-SSIM for Receiver 2, but suffers significant performance loss for Receiver 1.
Although TDMA considers the interests of both users, the BPG/JPEG2000 with TDMA is still inferior to the proposed DeepSCM in most of the simulated scenarios.
Specifically, for Receiver 1, the proposed scheme achieves a classification performance of 84$\%$, a PSNR performance of 21 dB and an MS-SSIM performance of 0.87 when $r=0.25$, outperforming the ``BPG+TDMA Capacity'' scheme by approximately 50$\%$ in classification, 6 dB in PSNR, and at least 0.35 in MS-SSIM.
For Receiver 2, when $r=0.04$, the classification performance of DeepSCM reaches 74$\%$, surpassing ``BPG+TDMA Capacity'' by up to 45$\%$.
DeepSCM also shows a PSNR advantage over it when $r<0.05$ and an MS-SSIM advantage when $r<0.1$, with up to a 0.11 improvement at $r = 0.04$.

Fig.~\ref{figure:rate 16} focuses on the cases where a 4QAM$\times$4QAM super-constellation is used.
Similar to the observations from Fig.~\ref{figure:rate 64}, our proposed scheme can achieve the upperbound performance of both Receiver 1 and 2 simultaneously on all three measures, and achieves a better balance between the performances of the two receivers than the ``BPG/JPEG2000+capacity'' baselines.

Finally, Fig.~\ref{example} displays visual examples of the image recovered by different schemes for Receiver 1 and Receiver 2.
Note that since BPG's performance is generally better than JPEG2000's, we omit the latter in this comparison.
It can be observed that the proposed scheme achieves relatively strong recovery performance for both receivers, while other schemes sacrifice the performance of one receiver for the benefit of the other.

\subsection{Performances versus Channel Disparity}

In this subsection, we compare the performance of DeepSCM with the ``CM trained jointly'' scheme and the ``BPG+LDPC'' scheme using TDMA at varying channel disparity levels, namely different SNR gaps between Receiver 1 and Receiver 2.
Specifically, we fix the channel SNR of Receiver 1 as $-5$ dB, and vary the channel SNR of Receiver 2 from 0 dB to 20 dB.
The transmission rate is set as 0.167 (512 channel uses).

Fig.~\ref{figure:snr}(a) and Fig.~\ref{figure:snr}(b) show the performance comparison for Receiver 1 and Receiver 2, respectively, at different channel disparity levels.
As a side note, we tune the hyperparameters $\lambda_1$ and $\lambda_2$ so that DeepSCM and ``CM trained jointly'' have the same performance in the classification task, and we compare their performances in the image recovery task.
Three important observations can be made.
First, comparing the proposed scheme and the ``BPG+LDPC'' scheme, we can see that for Receiver 1, DeepSCM outperforms ``BPG+LDPC'' by a significant margin; and for Receiver 2, DeepSCM outperforms the ``BPG+LDPC'' scheme in image recovery performance in low SNR regions ($<11$ dB) and constantly outperforms it in classification accuracy across all evaluated SNRs.
Moreover, the proposed scheme avoids the \textit{cliff effect} in the ``BPG+LDPC'' scheme, which is due to the failure of the channel code in adverse channel conditions with SNRs lower than the code's target SNR value.
Second, our scheme has an increasing PSNR performance advantage over ``CM trained jointly'' as the channel disparity increases.
For example, when using a 4QAM$\times$4QAM super-constellation, the performance of our scheme for Receiver 2 and that of the ``CM trained jointly'' scheme almost coincide when the channel disparity is lower than 15 dB.
When the channel disparity rises up to 25 dB, our scheme outperforms the ``CM trained jointly'' scheme by 0.25 dB.
Third, we can see that higher modulation order will further enlarge the performance gap between our scheme and the ``CM trained jointly'' scheme.
For example, when the channel disparity is 25 dB (the SNR of Receiver 2 is 20 dB), the PSNR performance gain of our scheme can reach 0.7 dB when a 4QAM$\times$16QAM super-constellation is used, which is 0.45 dB higher than the gain obtained when a 4QAM$\times$4QAM super-constellation is used. 

\begin{figure}[t]
     \centering
         \centering
         \includegraphics[width=0.49\textwidth]{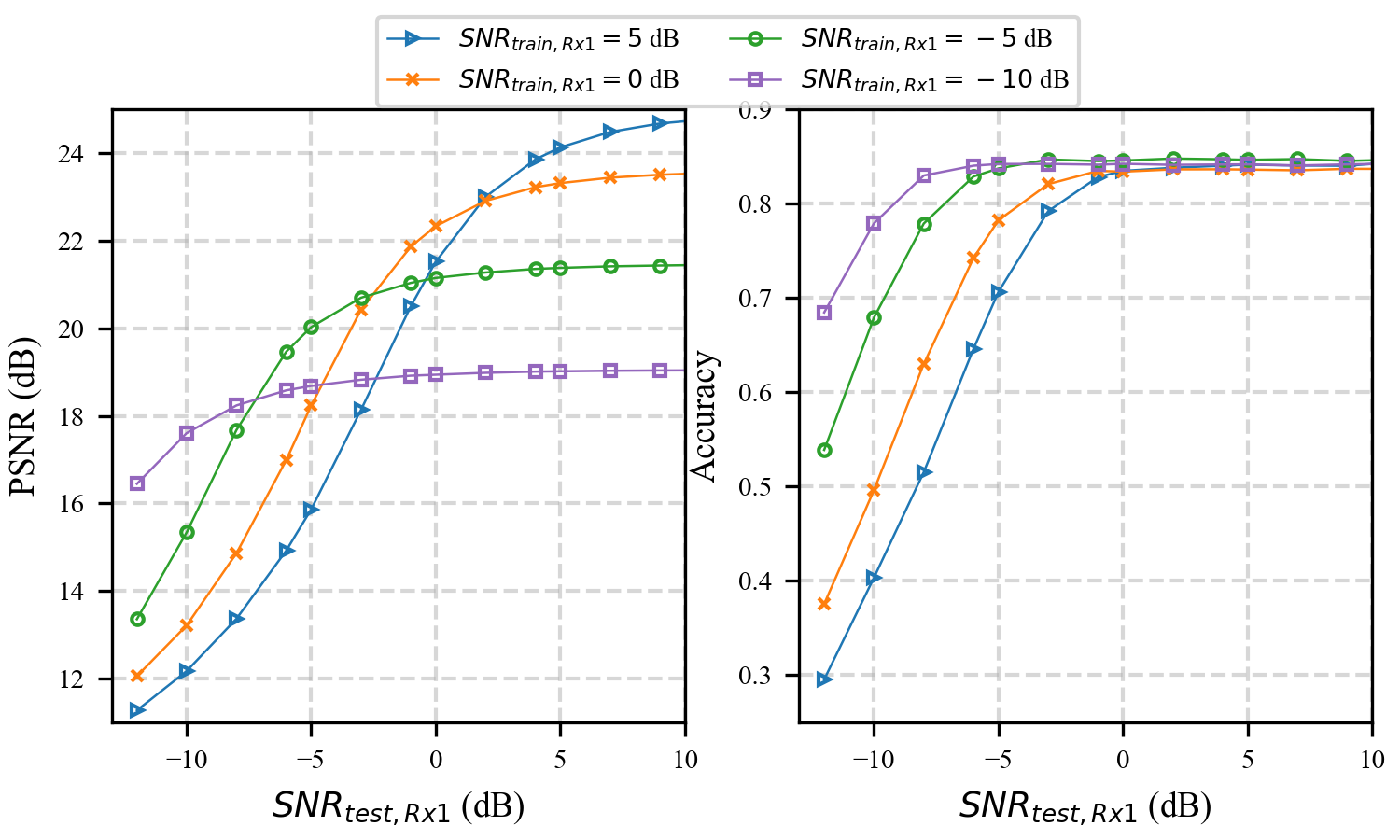}
    \caption{Performance of Receiver 1 trained at particular SNRs while tested across varying channel SNRs: (left) PSNR; (right) classification accuracy. Channel use is set as 512 with 4QAM$\times$16QAM. The channel SNR of Receiver 2 is set as 10 dB for both training and testing.}
    \label{figure:dynamic rx1}
    \vspace{-0.3cm}
\end{figure}

\begin{figure}[t]
     \centering
         \centering
         \includegraphics[width=0.49\textwidth]{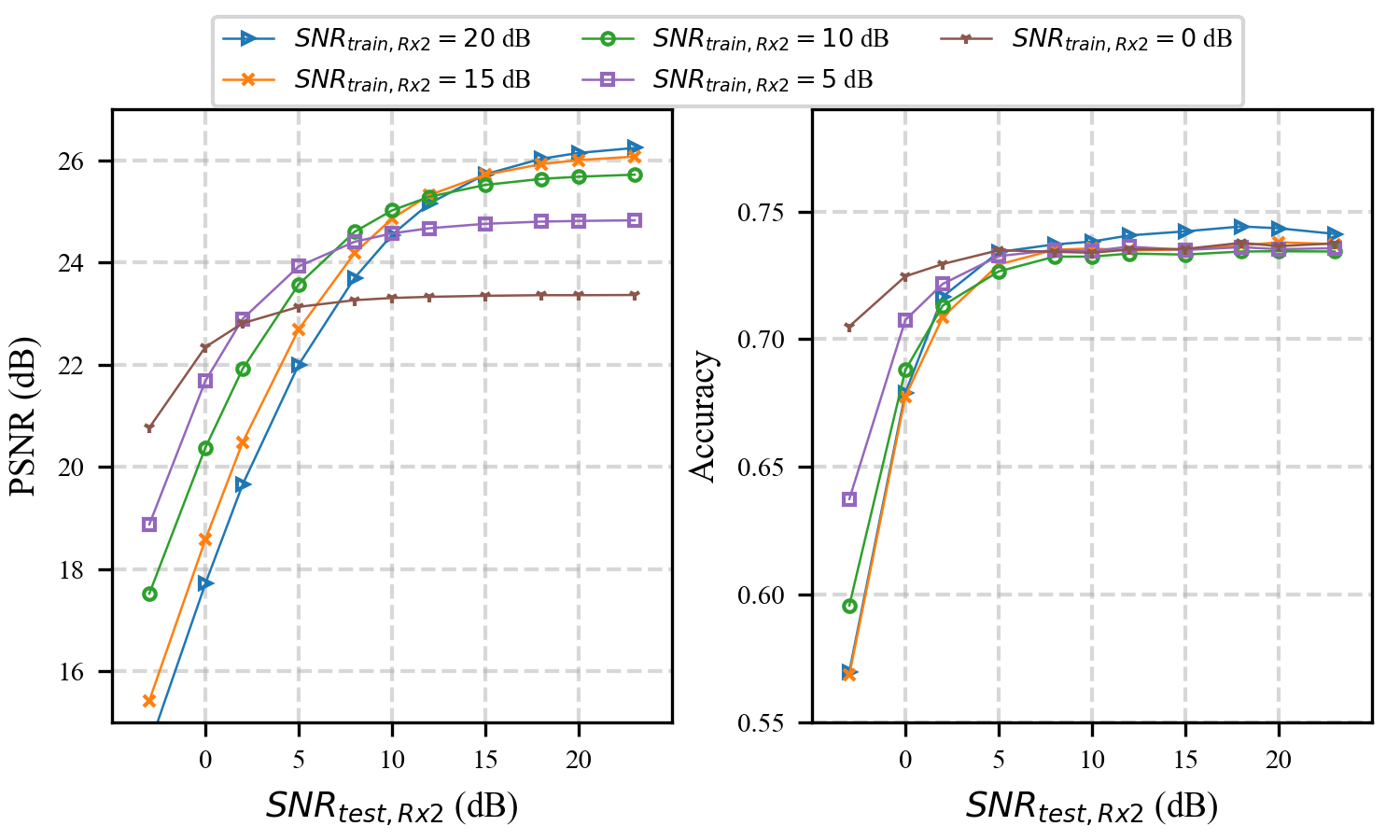}
    \caption{Performance of Receiver 2 trained at particular SNRs while tested across varying channel SNRs: (left) PSNR; (right) classification accuracy. Channel use is set as 512 with 4QAM$\times$16QAM. The channel SNR of Receiver 1 is set as -5 dB for both training and testing.}
    \label{figure:dynamic rx2}
    \vspace{-0.4cm}
\end{figure}

\begin{figure*}[t]
     \centering
     \begin{subfigure}[b]{0.49\textwidth}
         \centering
         \includegraphics[width=0.99\textwidth]{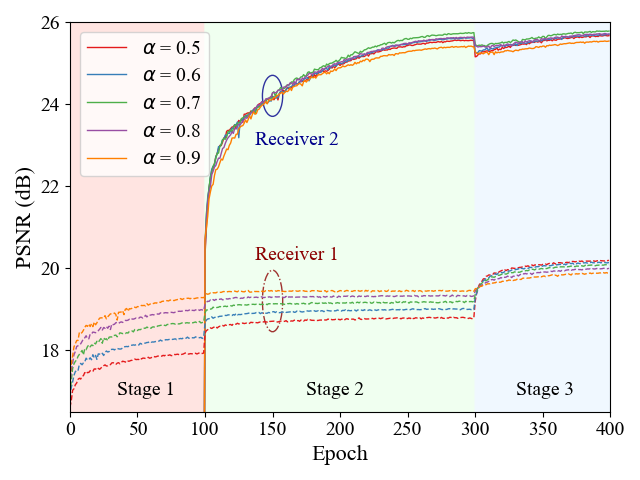}
         \caption{Convergence curve of image recovery.}
     \end{subfigure}
     \begin{subfigure}[b]{0.49\textwidth}
         \centering
         \includegraphics[width=0.99\textwidth]{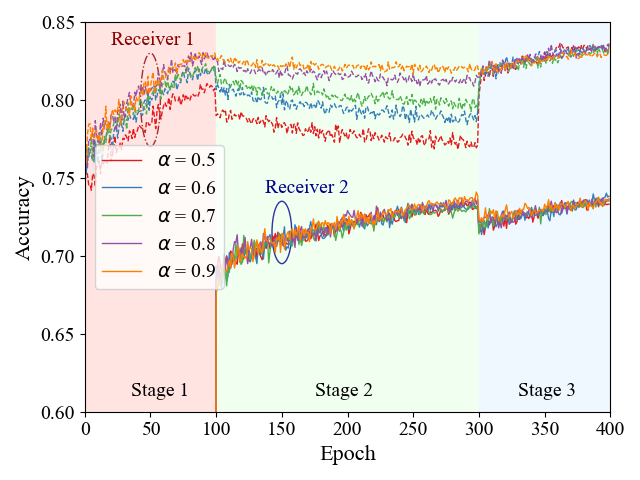}
         \caption{Convergence curve of classification accuracy.}
     \end{subfigure}
    \caption{Convergence of the DeepSCM scheme with varying PAF values.}
    \label{figure:a convergence}
    \vspace{-0.3cm}
\end{figure*}

\begin{figure*}[t]
     \centering
     \begin{subfigure}[b]{0.49\textwidth}
         \centering
         \includegraphics[width=0.99\textwidth]{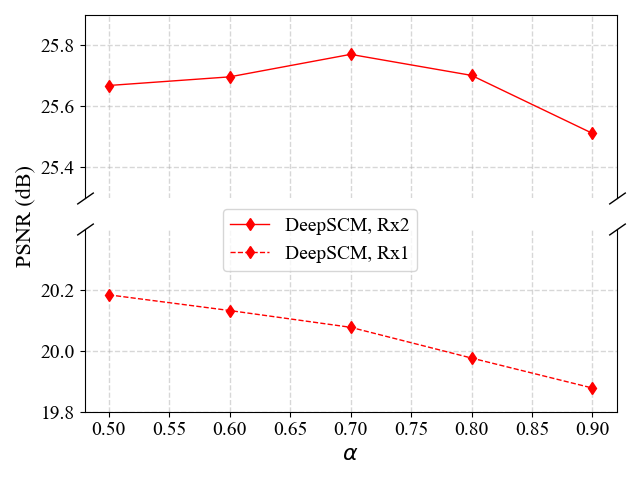}
         \caption{Image recovery \textit{vs.} $\alpha$.}
     \end{subfigure}
     \begin{subfigure}[b]{0.49\textwidth}
         \centering
         \includegraphics[width=0.99\textwidth]{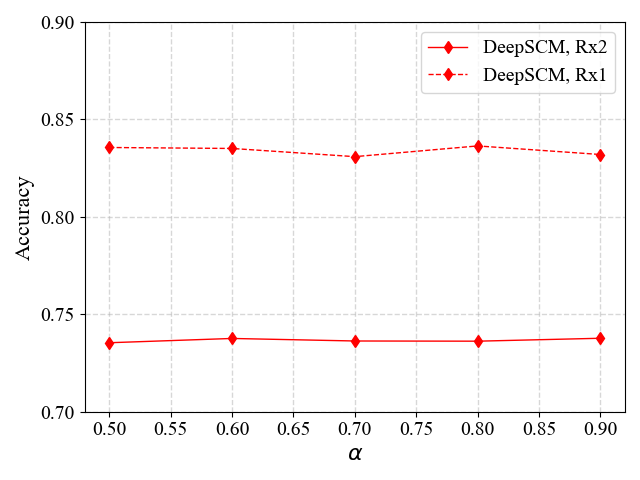}
         \caption{Classification accuracy \textit{vs.} $\alpha$.}
     \end{subfigure}
    \caption{Performance of the DeepSCM scheme with varying PAF values.}
    \label{figure:a performance}
    \vspace{-0.4cm}
\end{figure*}

\begin{figure*}
     \centering
     \begin{subfigure}[b]{0.19\textwidth}
         \centering
         \includegraphics[width=0.99\textwidth]{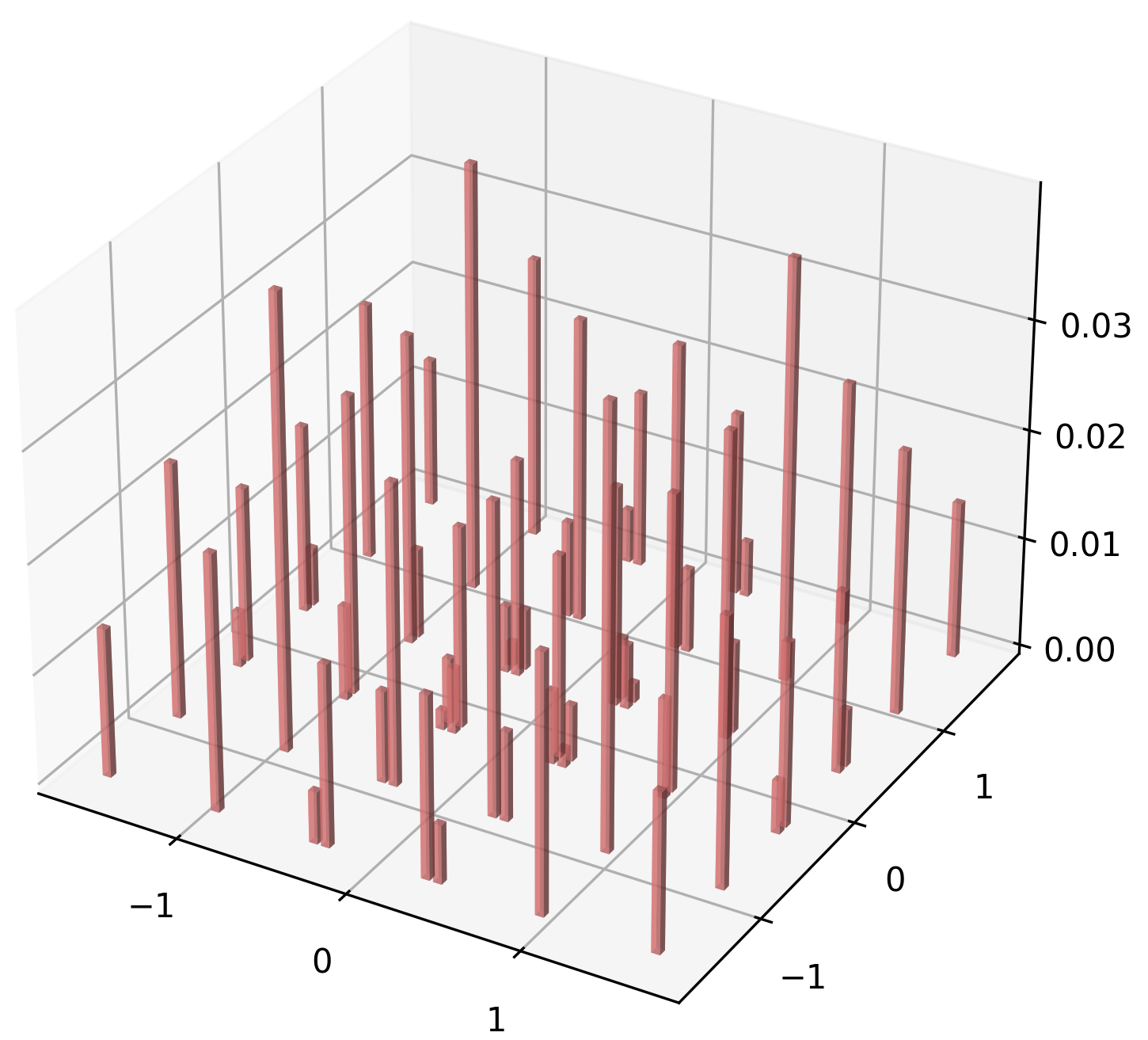}
         \caption{$a=0.5$.}
     \end{subfigure}
     \hfill
          \begin{subfigure}[b]{0.19\textwidth}
         \centering
         \includegraphics[width=0.99\textwidth]{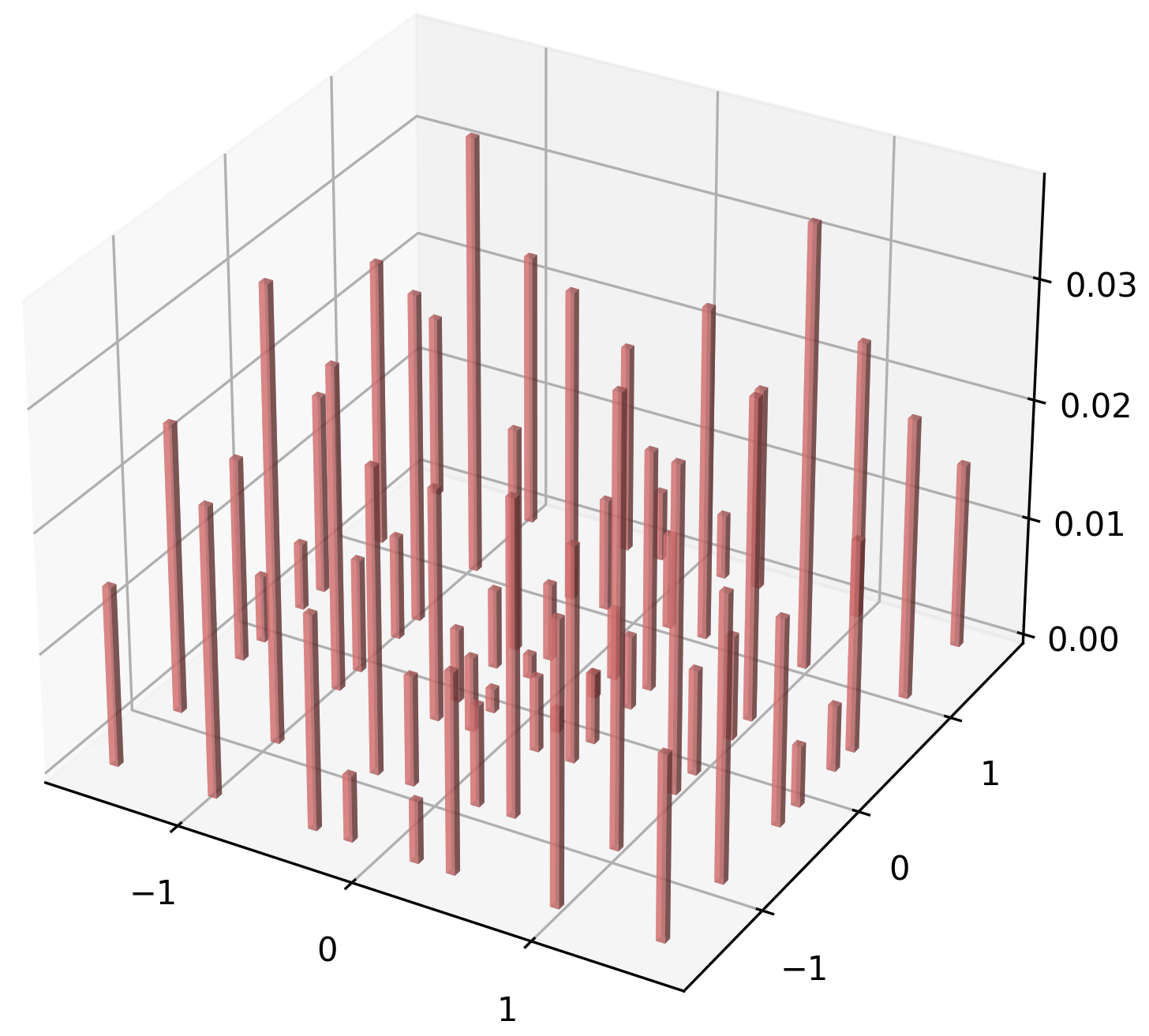}
         \caption{$a=0.6$.}
     \end{subfigure}
     \hfill
          \begin{subfigure}[b]{0.19\textwidth}
         \centering
         \includegraphics[width=0.99\textwidth]{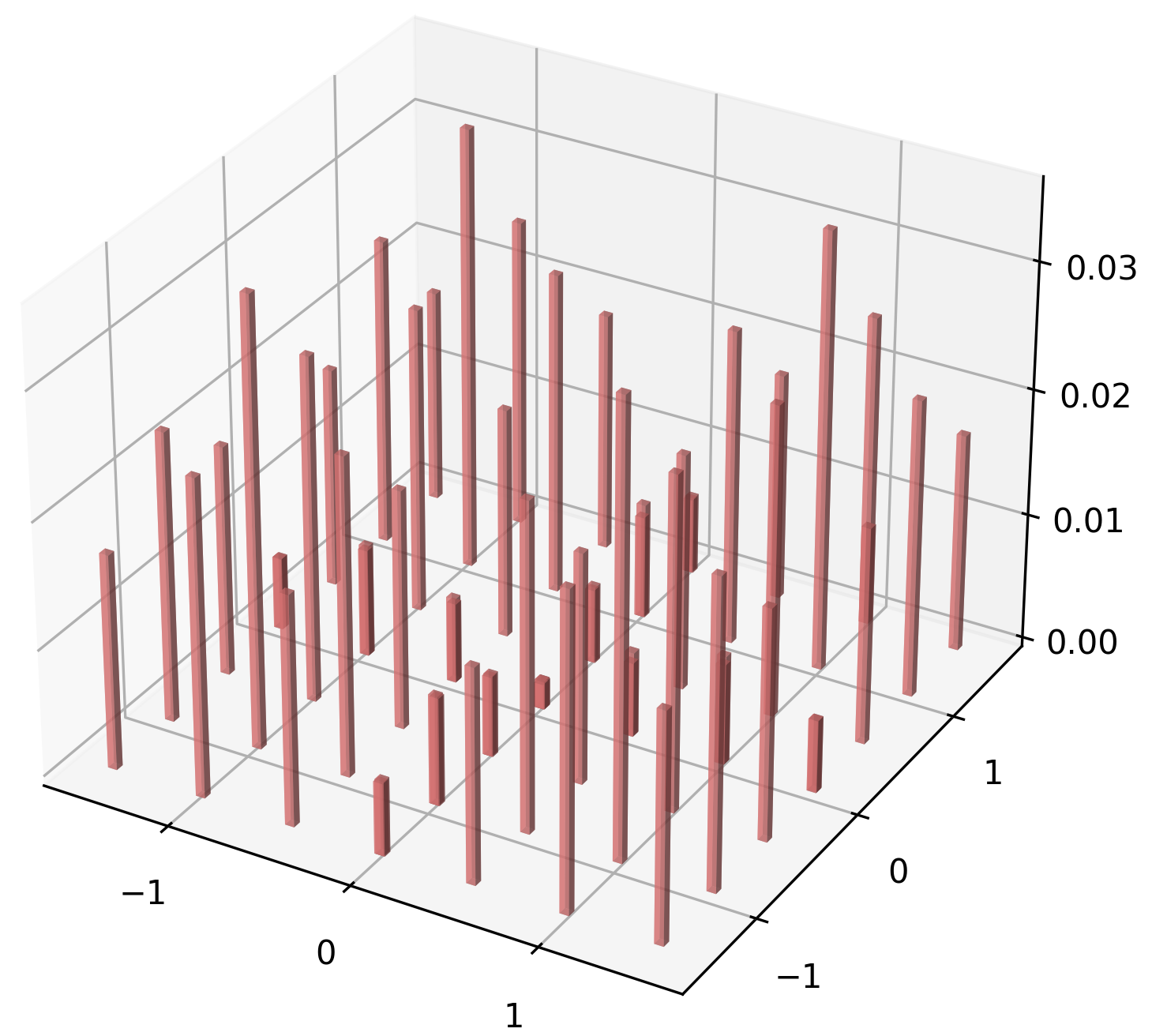}
         \caption{$a=0.7$.}
     \end{subfigure}
     \hfill
     \begin{subfigure}[b]{0.19\textwidth}
         \centering
         \includegraphics[width=0.99\textwidth]{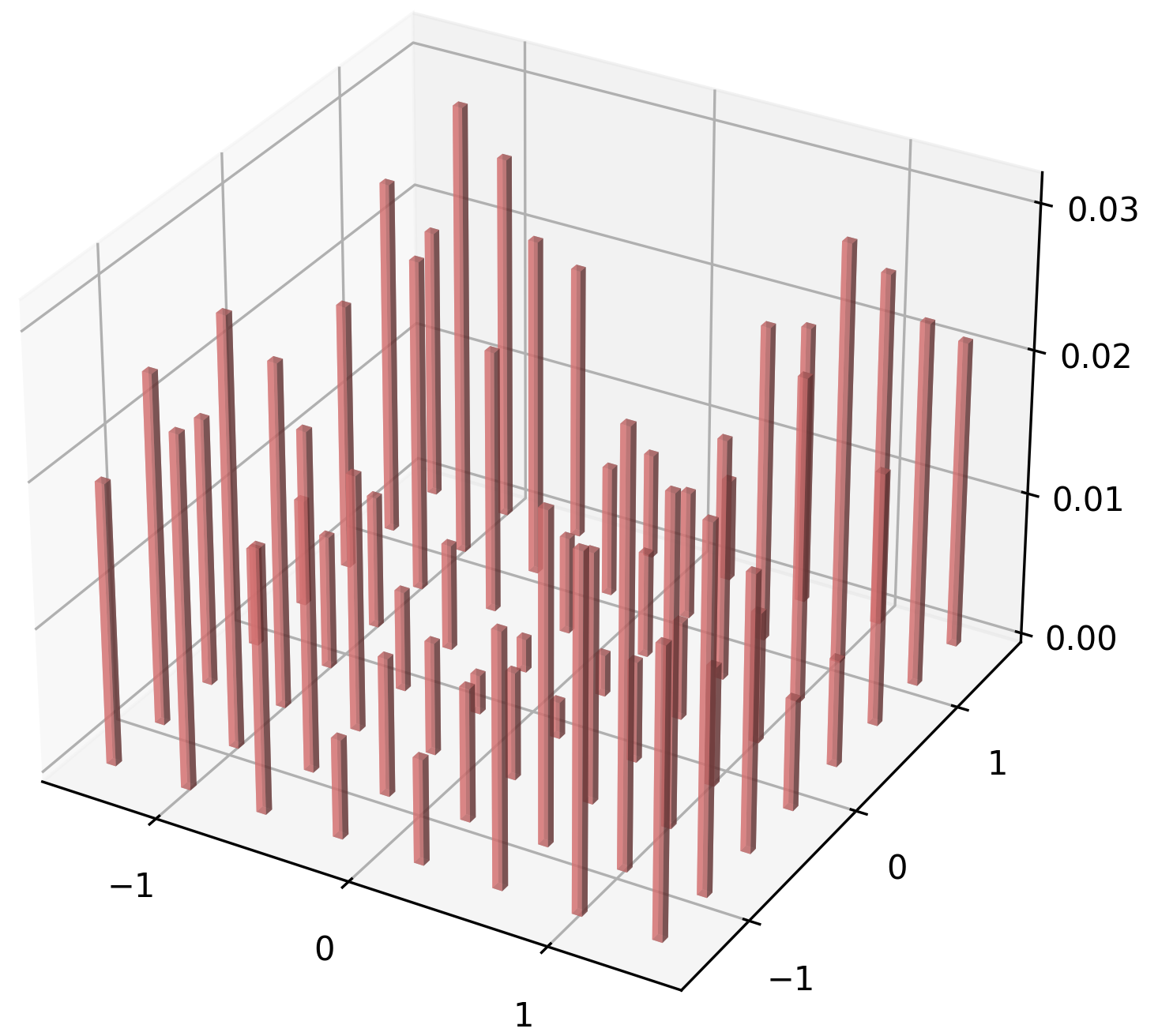}
         \caption{$a=0.8$.}
     \end{subfigure}
     \hfill
          \begin{subfigure}[b]{0.19\textwidth}
         \centering
         \includegraphics[width=0.99\textwidth]{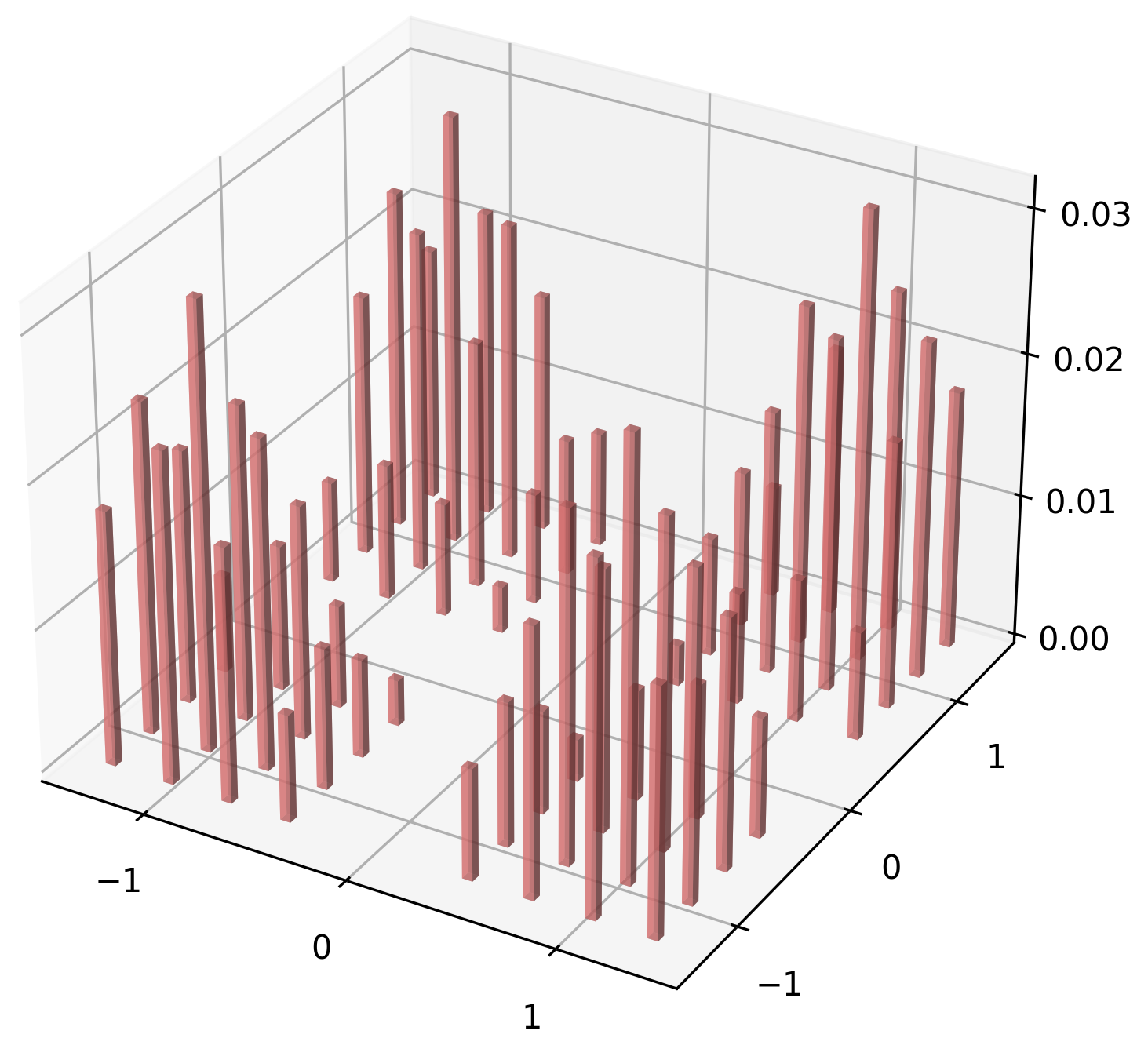}
         \caption{$a=0.9$.}
     \end{subfigure}
     \hfill
        \caption{Empirical distributions of the 4QAM$\times$16QAM super-constellation points with varying PAF values.
        }
        \label{empirical distribution}
        \vspace{-0.4cm}
\end{figure*}

In conclusion, the simulation results demonstrate that our scheme outperforms the ``CM trained jointly'' scheme, with its performance advantage increasing as the communication capability gap between the two receivers widens. 
Moreover, DeepSCM avoids the cliff effect experienced by the ``BPG+LDPC'' scheme and outperforms it in low SNR regions.

\subsection{Robustness to Variations in Channel Conditions}

This subsection investigates the robustness of the proposed scheme to variations in channel conditions.
To this end, the NNs in the system are trained at particular SNRs and tested over varying SNRs.
We denote the training SNR as $\text{SNR}_{\text{train}}$ and testing SNR as $\text{SNR}_{\text{test}}$.

Fig.~\ref{figure:dynamic rx1} and Fig.~\ref{figure:dynamic rx2} illustrate the performance of Receiver 1 and Receiver 2 respectively at varying $\text{SNR}_{\text{test}}$s.
For image recovery at both receivers, there is a graceful PSNR degradation when $\text{SNR}_{\text{test}}<\text{SNR}_{\text{train}}$,
and performance improves but gradually saturates when $\text{SNR}_{\text{test}}$ keeps increasing. 
For classification, the accuracy performance is much more robust to the variation of $\text{SNR}_{\text{test}}$ around $\text{SNR}_{\text{train}}$, compared to the image recovery.

Additionally, the system attains its peak performance at a specific SNR when optimized for it, and a bigger gap between $\text{SNR}_{\text{train}}$ and $\text{SNR}_{\text{test}}$ results in worse performance at this $\text{SNR}_{\text{test}}$.
This observation suggests a deployment strategy for the DeepSCM scheme in dynamic wireless conditions, \emph{i.e.}, storing multiple pretrained models trained with different channel SNRs at the transmitter and the receivers and then deploying the one with the training SNR closest to the current channel condition.

\subsection{Impact of PAF}

Power allocation between different receivers is an important issue in classical superposition code.
In this subsection, we investigate the impact of the hyperparameter PAF by varying the $\alpha$ in \eqref{superposition} from 0.5 to 0.9. 
The experiment is conducted using the 4QAM$\times$16QAM super-constellation. 
The channel SNR for Receiver 1 is set to -5 dB, while Receiver 2 has an SNR of 20 dB. 
We fix the transmission rate $r$ at 0.167 (equivalent to 512 channel uses). 
Intuitively, there should be a performance trade-off between the two receivers. 
Allocating more power to the inner constellation points will improve the performance of Receiver 1, and vice versa.

Fig.~\ref{figure:a convergence} shows the convergence of DeepSCM with different values of $\alpha$.
From Fig.~\ref{figure:a convergence}(a), we can indeed observe a trade-off in PSNR performance between the two receivers during the first two stages.
In Stage 1, larger values of $\alpha$ result in better performance for Receiver 1, which is attributed to the fact that the outer constellation points serve as additional noise.
In Stage 2, larger values of $\alpha$ (\emph{e.g.,} $\alpha=0.9$) lead to worse performance for Receiver 2.
This intuitive performance trade-off, however, disappears during the fine-tuning stage, where larger values of $\alpha$ also lead to decreased performance for Receiver 1.
This indicates that Receiver 1 can still decode the outer constellation to some extent due to the powerful adaptability of NNs to channel noises.
The influence of $\alpha$ on classification accuracy in Fig.~\ref{figure:a convergence}(b) is less obvious, yet similar observations can be made.

Fig.~\ref{figure:a performance} further compares the performance of the DeepSCM scheme with varying values of $\alpha$.
From the PSNR performance comparison in Fig.~\ref{figure:a performance}(a), it is shown that the performance of Receiver 1 decreases with increasing $\alpha$. This is because, during the fine-tuning stage, it can better decode the outer constellation with a smaller $\alpha$.
For Receiver 2, not only too large an $\alpha$ causes its performance degradation, but too small an $\alpha$ also leads to worse performance.
This is because with a smaller $\alpha$, the outer constellation points will overlay, making it harder for Receiver 2 to decode them.
The impact of PAF values on classification performance in Fig.~\ref{figure:a performance}(b) is observed to be less significant.
Finally, it's worth noting that these results differ from broadcasting independent messages to two receivers, given our hierarchical-source scenario. 
In the former case, increasing $\alpha$ improves the performance of one receiver while reducing that of the other.

Overall, these observations offer guidelines for determining the PAF: $\alpha$ should be chosen to prevent the overlay of outer constellation points, and the interval between outer constellation points should not be too small.
Therefore, it is best to choose moderate PAF values to achieve the overall best performance.
The observation that moderate PAF values are most reasonable significantly simplifies the selection of PAF.
We can simply set $\alpha$ such that the super-constellation forms rectangular QAM constellation, as in Section IV-B and Section IV-C.

In Figure~\ref{empirical distribution}, we illustrate the empirical distributions of the 4QAM$\times$16QAM super-constellations with varying PAF values. 
It can be noted that the outer constellation points overlap with a smaller $\alpha$. 
Moreover, it is evident that due to the poor channel condition of Receiver 1, constellation points with lower power are less likely to be generated.

\subsection{Complexity Analysis}

Table~\ref{flop} compares the computation complexity of the proposed DeepSCM with that of the CM scheme in terms of the number of floating point operations (FLOPs) and the number of model parameters. 
It can be seen that the receiver complexity of the two schemes is the same, but the transmitter complexity of DeepSCM is higher than CM. 
In particular, the transmitter of DeepSCM has 26.3$\%$ more parameters and almost doubles the FLOPs compared to the transmitter in the CM scheme.
This is expected as two JCM blocks are employed in the transmitter design of the proposed scheme.  

Table~\ref{delay} compares the inference delay of different schemes. 
The proposed DeepSCM scheme and the CM scheme exhibit similar average encoding and decoding time, and they are both faster than BPG, while JPEG2000 shows the shortest inference delay.
It is important to note that here the inference time of JPEG2000 and BPG refers to the source coding time only, and does not include the time for channel coding and task execution, \emph{i.e.}, classification in this work, which takes on average 7.2 ms.



\begin{table}[t]
\centering
\renewcommand{\arraystretch}{1.2}
\caption{The number of FLOPs and model parameters with $n=512$ channel uses.}
\footnotesize
\begin{tabular}{cc|c|c}
\hline
                                              &     & \textbf{FLOPs (G)} & \textbf{Params (M)} \\ \hline
\multicolumn{1}{c|}{\multirow{3}{*}{DeepSCM}} & Tx  & 4.3                & 102.7                                                                                  \\ \cline{2-4} 
\multicolumn{1}{c|}{}                         & Rx1 & 1.5                & 24.7                                                                                   \\ \cline{2-4} 
\multicolumn{1}{c|}{}                         & Rx2 & 1.5                & 24.9                                                                                  \\ \hline
\multicolumn{1}{c|}{\multirow{3}{*}{CM}}      & Tx  & 2.2                & 81.3                                                                                   \\ \cline{2-4} 
\multicolumn{1}{c|}{}                         & Rx1 & 1.5                & 24.7                                                                                  \\ \cline{2-4} 
\multicolumn{1}{c|}{}                         & Rx2 & 1.5                & 24.9                                                                                   \\ \hline
\end{tabular}
\label{flop}
\end{table}

\begin{table}[t]
\centering
\renewcommand{\arraystretch}{1.2}
\caption{Inference delay comparison.}
\footnotesize
\begin{tabular}{c|c|c}
\hline
         & \textbf{Encoding Time (ms)} & \textbf{Decoding Time (ms)} \\ \hline
SCM      & 16.9               & 11.0                \\ \hline
CM       & 16.3               & 11.0                 \\ \hline
JPEG2000 & 3.8                & 0.6                  \\ \hline
BPG      & 28.3             & 18.5               \\ \hline
\end{tabular}
\label{delay}
\vspace{-0.4cm}
\end{table}

\section{Conclusion}

This paper proposes a new framework for digital semantic communications over degraded AWGN broadcast channels, namely the DeepSCM scheme.
In this scheme, the semantic features intended for different receivers are encoded into a basic EFV and its successive refinement vector, which are then associated with different layers of a super-constellation.
To minimize redundancy in broadcasting, an LMMSE decorrelator is developed to ensure that these two vectors are nearly uncorrelated with each other.
This superposition code structure can accommodate the communication requirements of different receivers with diverse channel conditions.
The proposed scheme is especially effective in scenarios with large channel disparity and high modulation order. 
This work not only provides an efficient way to conduct semantic broadcasting, but also showcases a promising approach of combining theoretical coding schemes with NN-based coding method. 
\begin{appendices}

\section{Proof of Proposition \ref{upperbound}}

The proof of Proposition \ref{upperbound} follows a straightforward idea of upper-bounding the differential entropy of $\mathbf{R}$ step by step. 
The details are as follows.
\begin{align}
    h(\mathbf{R})
    &\overset{(a)}{\le} \frac{1}{2}\log\left\lbrace(2\pi e)^{2n}|\textit{Var}[\mathbf{R}]|\right\rbrace\nonumber\\
    &\overset{(b)}{\le} \frac{1}{2}\log\left\lbrace(2\pi e)^{2n} \textstyle \prod_{i=1}^{2n} \textit{Var}[R_i]\right\rbrace \nonumber\\
    &\overset{(c)}{\le}\frac{1}{2}\log\left\lbrace(2\pi e)^{2n}(\frac{\sum_{i=1}^{2n} \textit{Var}[R_i]}{2n})^{2n}\right\rbrace\nonumber\\
    &\overset{(d)}{\le}n\log\left\lbrace \frac{\pi e}{n}\sum_{i=1}^{2n} \mathbf{E}[R_{i}^ 2]\right\rbrace\nonumber\\
    &=n\log \left\lbrace\frac{\pi e}{n} \mathbf{E}[||\mathbf{R}||^2_2]\right\rbrace,
\end{align}
where $R_i$ denotes the $i$th entry of $\mathbf{R}$, $|\textit{Var}[\mathbf{R}]|$ denotes the determinant of $\textit{Var}[\mathbf{R}]$, $(a)$ follows from the fact that Gaussian distribution maximizes entropy under a given covariance matrix, $(b)$ follows from Hadamard inequality since covariance matrices are always positive semidefinite, $(c)$ follows from the inequality of arithmetic and geometric means, and $(d)$ follows from the definition of variance that $\textit{Var}[R_i]=\mathbf{E}[R_{i}^2]-\mathbf{E}^2[R_i]$.

We have equality in $(a)$ if $\mathbf{R}$ is Gaussian distributed; we have equality in $(b)$ if the elements of $\mathbf{R}$ are mutually uncorrelated; the equality in $(c)$ holds if $\textit{Var}[R_1]=\textit{Var}[R_2]=...=\textit{Var}[R_{2n}]$ according to the inequality of arithmetic and geometric means; and the equality in $(d)$ holds if all elements in $\mathbf{R}$ have zero mean. 
Combining these facts together suggests the equality holds if each element of $\mathbf{R}$ follows an i.i.d. Gaussian distribution with zero mean.

\end{appendices}

\bibliographystyle{IEEEtran}
\bibliography{scm_journal}{}

\begin{thebibliography}{10}
\providecommand{\url}[1]{#1}
\csname url@samestyle\endcsname
\providecommand{\newblock}{\relax}
\providecommand{\bibinfo}[2]{#2}
\providecommand{\BIBentrySTDinterwordspacing}{\spaceskip=0pt\relax}
\providecommand{\BIBentryALTinterwordstretchfactor}{4}
\providecommand{\BIBentryALTinterwordspacing}{\spaceskip=\fontdimen2\font plus
\BIBentryALTinterwordstretchfactor\fontdimen3\font minus
  \fontdimen4\font\relax}
\providecommand{\BIBforeignlanguage}[2]{{%
\expandafter\ifx\csname l@#1\endcsname\relax
\typeout{** WARNING: IEEEtran.bst: No hyphenation pattern has been}%
\typeout{** loaded for the language `#1'. Using the pattern for}%
\typeout{** the default language instead.}%
\else
\language=\csname l@#1\endcsname
\fi
#2}}
\providecommand{\BIBdecl}{\relax}
\BIBdecl

\bibitem{confVersion}
Y.~Bo, S.~Shao, and M.~Tao, ``A superposition code approach for digital
  semantic communications over broadcast channels,'' in \emph{GLOBECOM
  2023-2023 IEEE Global Communications Conference}.\hskip 1em plus 0.5em minus
  0.4em\relax IEEE, 2023, pp. 1586--1591.

\bibitem{survey:beyond}
D.~Gündüz, Z.~Qin, I.~E. Aguerri, H.~S. Dhillon, Z.~Yang, A.~Yener, K.~K.
  Wong, and C.-B. Chae, ``Beyond transmitting bits: Context, semantics, and
  task-oriented communications,'' \emph{IEEE Journal on Selected Areas in
  Communications}, vol.~41, no.~1, pp. 5--41, 2023.

\bibitem{survey:view}
Q.~Lan, D.~Wen, Z.~Zhang, Q.~Zeng, X.~Chen, P.~Popovski, and K.~Huang, ``What
  is semantic communication? a view on conveying meaning in the era of machine
  intelligence,'' \emph{Journal of Communications and Information Networks},
  vol.~6, no.~4, pp. 336--371, 2021.

\bibitem{survey:overview}
X.~Luo, H.-H. Chen, and Q.~Guo, ``Semantic communications: Overview, open
  issues, and future research directions,'' \emph{IEEE Wireless
  Communications}, vol.~29, no.~1, pp. 210--219, 2022.

\bibitem{survey:opportunity}
M.~Sana and E.~C. Strinati, ``Learning semantics: An opportunity for effective
  6g communications,'' in \emph{2022 IEEE 19th Annual Consumer Communications
  And Networking Conference (CCNC)}, 2022, pp. 631--636.

\bibitem{semantic:speech}
Z.~Weng and Z.~Qin, ``Semantic communication systems for speech transmission,''
  \emph{IEEE Journal on Selected Areas in Communications}, vol.~39, no.~8, pp.
  2434--2444, 2021.

\bibitem{semantic:text}
H.~Xie, Z.~Qin, G.~Y. Li, and B.-H. Juang, ``Deep learning enabled semantic
  communication systems,'' \emph{IEEE Transactions on Signal Processing},
  vol.~69, pp. 2663--2675, 2021.

\bibitem{semantic:text2}
Q.~Zhou, R.~Li, Z.~Zhao, C.~Peng, and H.~Zhang, ``Semantic communication with
  adaptive universal transformer,'' \emph{IEEE Wireless Communications
  Letters}, vol.~11, no.~3, pp. 453--457, 2022.

\bibitem{semantic:text3}
H.~Xie and Z.~Qin, ``A lite distributed semantic communication system for
  internet of things,'' \emph{IEEE Journal on Selected Areas in
  Communications}, vol.~39, no.~1, pp. 142--153, 2021.

\bibitem{semantic:image}
J.~Shao, Y.~Mao, and J.~Zhang, ``Learning task-oriented communication for edge
  inference: An information bottleneck approach,'' \emph{IEEE Journal on
  Selected Areas in Communications}, vol.~40, no.~1, pp. 197--211, 2022.

\bibitem{semantic:image2}
K.~Liu, D.~Liu, L.~Li, N.~Yan, and H.~Li, ``Semantics-to-signal scalable image
  compression with learned revertible representations,'' \emph{International
  Journal of Computer Vision}, vol. 129, no.~9, pp. 2605--2621, 2021.

\bibitem{semantic:image3}
H.~Zhang, S.~Shao, M.~Tao, X.~Bi, and K.~B. Letaief, ``Deep learning-enabled
  semantic communication systems with task-unaware transmitter and dynamic
  data,'' \emph{IEEE Journal on Selected Areas in Communications}, vol.~41,
  no.~1, pp. 170--185, 2023.

\bibitem{gunduz2019jscc1}
E.~Bourtsoulatze, D.~Burth~Kurka, and D.~Gündüz, ``Deep joint source-channel
  coding for wireless image transmission,'' \emph{IEEE Transactions on
  Cognitive Communications and Networking}, vol.~5, no.~3, pp. 567--579, 2019.

\bibitem{semantic:video}
S.~Wang, J.~Dai, Z.~Liang, K.~Niu, Z.~Si, C.~Dong, X.~Qin, and P.~Zhang,
  ``Wireless deep video semantic transmission,'' \emph{IEEE Journal on Selected
  Areas in Communications}, vol.~41, no.~1, pp. 214--229, 2023.

\bibitem{semantic:video2}
P.~Jiang, C.-K. Wen, S.~Jin, and G.~Y. Li, ``Wireless semantic communications
  for video conferencing,'' \emph{IEEE Journal on Selected Areas in
  Communications}, vol.~41, no.~1, pp. 230--244, 2023.

\bibitem{mac}
H.~Xie, Z.~Qin, and G.~Y. Li, ``Task-oriented multi-user semantic
  communications for {VQA},'' \emph{IEEE Wireless Communications Letters},
  vol.~11, no.~3, pp. 553--557, 2022.

\bibitem{objectdetection}
D.~Huang, F.~Gao, X.~Tao, Q.~Du, and J.~Lu, ``Toward semantic communications:
  Deep learning-based image semantic coding,'' \emph{IEEE Journal on Selected
  Areas in Communications}, vol.~41, no.~1, pp. 55--71, 2023.

\bibitem{resnet}
K.~He, X.~Zhang, S.~Ren, and J.~Sun, ``Deep residual learning for image
  recognition,'' in \emph{2016 IEEE Conference on Computer Vision and Pattern
  Recognition (CVPR)}, 2016, pp. 770--778.

\bibitem{vaswani2017attention}
A.~Vaswani, N.~Shazeer, N.~Parmar, J.~Uszkoreit, L.~Jones, A.~N. Gomez,
  {\L}.~Kaiser, and I.~Polosukhin, ``Attention is all you need,'' in
  \emph{Advances in neural information processing systems}, vol.~30, 2017, pp.
  5998--6008.

\bibitem{devlin2018bert}
J.~Devlin, M.-W. Chang, K.~Lee, and K.~Toutanova, ``{BERT}: Pre-training of
  deep bidirectional transformers for language understanding,'' \emph{arXiv
  preprint arXiv:1810.04805}, 2018.

\bibitem{one2many}
H.~Hu, X.~Zhu, F.~Zhou, W.~Wu, R.~Q. Hu, and H.~Zhu, ``One-to-many semantic
  communication systems: Design, implementation, performance evaluation,''
  \emph{IEEE Communications Letters}, vol.~26, no.~12, pp. 2959--2963, 2022.

\bibitem{vectorquantization}
Q.~Fu, H.~Xie, Z.~Qin, G.~Slabaugh, and X.~Tao, ``Vector quantized semantic
  communication system,'' \emph{IEEE Wireless Communications Letters}, vol.~12,
  no.~6, pp. 982--986, 2023.

\bibitem{xie2023robust}
S.~Xie, S.~Ma, M.~Ding, Y.~Shi, M.~Tang, and Y.~Wu, ``Robust information
  bottleneck for task-oriented communication with digital modulation,''
  \emph{IEEE Journal on Selected Areas in Communications}, 2023.

\bibitem{jaccq}
T.-Y. Tung, D.~B. Kurka, M.~Jankowski, and D.~Gündüz, ``{DeepJSCC-Q}:
  Constellation constrained deep joint source-channel coding,'' \emph{IEEE
  Journal on Selected Areas in Information Theory}, vol.~3, no.~4, pp.
  720--731, 2022.

\bibitem{jcm}
Y.~Bo, Y.~Duan, S.~Shao, and M.~Tao, ``Learning based joint coding-modulation
  for digital semantic communication systems,'' in \emph{2022 14th
  International Conference on Wireless Communications and Signal Processing
  (WCSP)}, 2022, pp. 1--6.

\bibitem{bo2023joint}
------, ``Joint coding-modulation for digital semantic communications via
  variational autoencoder,'' \emph{IEEE Trans. Commun.}, pp. 1--15, 2024,
  \textit{Early Access}, doi: 10.1109/TCOMM.2024.3386577.

\bibitem{distributedJSCC}
S.~Wang, K.~Yang, J.~Dai, and K.~Niu, ``Distributed image transmission using
  deep joint source-channel coding,'' in \emph{ICASSP 2022 - 2022 IEEE
  International Conference on Acoustics, Speech and Signal Processing
  (ICASSP)}, 2022, pp. 5208--5212.

\bibitem{yilmaz2023distributed}
S.~F. Yilmaz, C.~Karamanl{\i}, and D.~G{\"u}nd{\"u}z, ``Distributed deep joint
  source-channel coding over a multiple access channel,'' in \emph{ICC
  2023-IEEE International Conference on Communications}.\hskip 1em plus 0.5em
  minus 0.4em\relax IEEE, 2023, pp. 1400--1405.

\bibitem{lin2023channel}
L.~Lin, W.~Xu, F.~Wang, Y.~Zhang, W.~Zhang, and P.~Zhang,
  ``Channel-transferable semantic communications for multi-user ofdm-noma
  systems,'' \emph{IEEE Wireless Communications Letters}, 2023.

\bibitem{cooparative}
Y.~Zhang, W.~Xu, H.~Gao, and F.~Wang, ``Multi-user semantic communications for
  cooperative object identification,'' in \emph{2022 IEEE International
  Conference on Communications Workshops (ICC Workshops)}, 2022, pp. 157--162.

\bibitem{qin:multiuser}
H.~Xie, Z.~Qin, X.~Tao, and K.~B. Letaief, ``Task-oriented multi-user semantic
  communications,'' \emph{IEEE Journal on Selected Areas in Communications},
  vol.~40, no.~9, pp. 2584--2597, 2022.

\bibitem{mdma}
P.~Zhang, X.~Xu, C.~Dong, K.~Niu, H.~Liang, Z.~Liang, X.~Qin, M.~Sun, H.~Chen,
  N.~Ma \emph{et~al.}, ``Model division multiple access for semantic
  communications,'' \emph{Frontiers of Information Technology \& Electronic
  Engineering}, pp. 1--12, 2023.

\bibitem{ma2023features}
S.~Ma, Z.~Zhang, Y.~Wu, H.~Li, G.~Shi, D.~Gao, Y.~Shi, S.~Li, and N.~Al-Dhahir,
  ``Features disentangled semantic broadcast communication networks,''
  \emph{IEEE Transactions on Wireless Communications}, pp. 1--1, 2023.

\bibitem{wu2023fusion}
T.~Wu, Z.~Chen, M.~Tao, B.~Xia, and W.~Zhang, ``Fusion-based multi-user
  semantic communications for wireless image transmission over degraded
  broadcast channels,'' in \emph{GLOBECOM 2023-2023 IEEE Global Communications
  Conference}.\hskip 1em plus 0.5em minus 0.4em\relax IEEE, 2023, pp.
  7623--7628.

\bibitem{li2023non}
W.~Li, H.~Liang, C.~Dong, X.~Xu, P.~Zhang, and K.~Liu, ``Non-orthogonal
  multiple access enhanced multi-user semantic communication,'' \emph{IEEE
  Transactions on Cognitive Communications and Networking}, vol.~9, no.~6, pp.
  1438--1453, 2023.

\bibitem{lu2023self}
Z.~Lu, R.~Li, M.~Lei, C.~Wang, Z.~Zhao, and H.~Zhang, ``Self-critical alternate
  learning based semantic broadcast communication,'' \emph{arXiv preprint
  arXiv:2312.01423}, 2023.

\bibitem{cover:broadcast}
T.~Cover, ``Broadcast channels,'' \emph{IEEE Transactions on Information
  Theory}, vol.~18, no.~1, pp. 2--14, 1972.

\bibitem{image1}
K.~Liu, D.~Liu, L.~Li, N.~Yan, and H.~Li, ``Semantics-to-signal scalable image
  compression with learned revertible representations,'' \emph{International
  Journal of Computer Vision}, vol. 129, no.~9, pp. 2605--2621, 2021.

\bibitem{jang2016softgumbel}
E.~Jang, S.~Gu, and B.~Poole, ``Categorical reparameterization with
  gumbel-softmax,'' \emph{arXiv:1611.01144}, 2016.

\bibitem{kay1993fundamentals}
S.~M. Kay, \emph{Fundamentals of statistical signal processing: estimation
  theory}.\hskip 1em plus 0.5em minus 0.4em\relax Prentice-Hall, Inc., 1993.

\bibitem{krizhevsky2009cifar}
A.~Krizhevsky, G.~Hinton \emph{et~al.}, ``Learning multiple layers of features
  from tiny images,'' \emph{University of Toronto, Technical Report}, 2009.

\bibitem{spinal}
H.~M.~D. Kabir, M.~Abdar, A.~Khosravi, S.~M.~J. Jalali, A.~F. Atiya,
  S.~Nahavandi, and D.~Srinivasan, ``Spinalnet: Deep neural network with
  gradual input,'' \emph{IEEE Transactions on Artificial Intelligence}, pp.
  1--13, 2022.

\bibitem{wang2003multiscale}
Z.~Wang, E.~P. Simoncelli, and A.~C. Bovik, ``Multiscale structural similarity
  for image quality assessment,'' in \emph{The Thrity-Seventh Asilomar
  Conference on Signals, Systems $\&$ Computers, 2003}, vol.~2.\hskip 1em plus
  0.5em minus 0.4em\relax Ieee, 2003, pp. 1398--1402.

\end{thebibliography}

\end{document}